\documentclass[%
 reprint,
 superscriptaddress,
 nofootinbib,
 amsmath,amssymb,
 aps,
 pra,
]{revtex4-1}

\usepackage[pdftex]{graphicx}
\usepackage[pdftex,
            colorlinks=true,
            citecolor=blue,
            linkcolor=blue]{hyperref}
\usepackage{newtxtext, newtxmath}
\usepackage{bm}
\usepackage{braket, mathtools}
\usepackage{tabularx, multirow}
\newcolumntype{C}[1]{>{\centering\arraybackslash}p{#1}}
\newcolumntype{L}[1]{>{\raggedright\arraybackslash}p{#1}}
\newcolumntype{R}[1]{>{\raggedleft\arraybackslash}p{#1}}

\DeclareMathOperator{\Tr}{Tr}
\DeclareMathOperator{\sgn}{sgn}

\begin{document}

\title{Superconductivity induced by fluctuations of momentum-based multipoles}

\author{Shuntaro Sumita}
\email[]{shuntaro.sumita@riken.jp}
\affiliation{%
 Condensed Matter Theory Laboratory, RIKEN CPR, Wako, Saitama 351-0198, Japan
}%
\affiliation{%
 Department of Physics, Kyoto University, Kyoto 606-8502, Japan
}%

\author{Youichi Yanase}
\affiliation{%
 Department of Physics, Kyoto University, Kyoto 606-8502, Japan
}%
\affiliation{%
 Institute for Molecular Science, 38 Nishigo-Naka, Myodaiji, Okazaki, Aichi 444-8585, Japan
}%

\date{\today}

\begin{abstract}
Recent studies of unconventional superconductivity have focused on charge or spin fluctuation, instead of electron-phonon coupling, as an origin of attractive interaction between electrons.
On the other hand, a multipole order, which represents electrons' degrees of freedom in strongly correlated and spin-orbit-coupled systems, has recently been attracting much attention.
Stimulated by this background, we investigate \textit{multipole-fluctuation-mediated superconductivity}, which proposes a new pairing mechanism of unconventional superconductivity.
Indeed, previous works have shown spin-triplet superconductivity induced by fluctuations of odd-parity electric multipole orders in isotropic systems.
In this study, we establish a general formulation of the multipole-fluctuation-mediated superconductivity for all multipole symmetries, in both isotropic and crystalline systems.
As a result, we reveal various anisotropic pairings induced by odd-parity and/or higher-order multipole fluctuations, which are beyond the ordinary charge or spin fluctuations.
Topological superconductivity due to the mechanism is also discussed.
Based on the obtained results, we discuss unconventional superconductivity in doped SrTiO$_3$, PrTi$_2$Al$_{20}$, Li$_2$(Pd,~Pt)$_3$B, and magnetic multipole metals.
\end{abstract}

\maketitle

\section{Introduction}
Searching for unconventional superconductors is one of the central issues in recent condensed matter physics.
While the conventional BCS superconductivity is mediated by an electron-phonon coupling, recent studies have elucidated charge- or spin-fluctuation-induced unconventional superconductivity, e.g., $d$-wave superconductivity in high-$T_{\text{c}}$ cuprate superconductors and CeCoIn$_5$ by antiferromagnetic spin fluctuations~\cite{Miyake1986, Moriya2000, Yanase2003}, spin-triplet superconductivity in UCoGe and URhGe by ferromagnetic spin fluctuations~\cite{Hattori2014, Aoki2019_review} analogous to superfluid $^3$He~\cite{Nakajima1973}, and $s_{++}$-wave superconductivity in Fe-based superconductors by charge (orbital) fluctuations~\cite{Yanagi2010, Kontani2010}.
Thus, the correlation between superconductivity and other electric or magnetic orders has attracted much attention, especially in the field of strongly correlated electron systems.

In the field, on the other hand, recent theoretical and experimental studies have vigorously reinterpreted various electric or magnetic orders, as well as charge or spin, in the context of \textit{multipoles}~\cite{Kuramoto2009_review, Spaldin2008, Spaldin2013, Yanase2014, Hitomi2014, Hayami2014_1, Hayami2014_2, Fu2015, Hitomi2016, Sumita2016, Sumita2017, Suzuki2017, Suzuki2019, Higo2018, Watanabe2018, Hayami2018, Saito2018, Shitade2018, Shitade2019, Hitomi2019, Sakai2011, Ito2011, Sato2012, Sakai2012, Matsubayashi2012, Tsujimoto2014, Haule2009, Kusunose2011, Ikeda2012, Koga2006}.
Therefore, attempts to extend the above-mentioned mechanism of superconductivity to \textit{multipole-fluctuation-mediated superconductivity} are attracting interest.
A topic of interest is higher-order multipoles in superconductors.
For example, an electric quadrupole order may be closely related to superconductivity in Pr\textit{Tr}$_2$Al$_{20}$ (\textit{Tr}~$=$~Ti, V)~\cite{Sakai2011, Ito2011, Sato2012, Sakai2012, Matsubayashi2012, Tsujimoto2014} and PrIr$_2$Zn$_{20}$~\cite{Onimaru2010, Onimaru2011}; an electric hexadecapole~\cite{Haule2009, Kusunose2011} or a magnetic dotriacontapole order~\cite{Ikeda2012} in URu$_2$Si$_2$ and various multipoles in PrOs$_4$Sb$_{12}$~\cite{Koga2006} have also been intensively discussed.
Another topic is odd-parity multipoles; recent theoretical studies have pointed out that odd-parity multipoles invoke unconventional superconductivity not only in the coexisting state~\cite{Sumita2016, Sumita2017, Kanasugi2018, Kanasugi2019} but also in the disordered state due to the multipole fluctuations~\cite{Kozii2015, Kozii2019, Lee2020, Gastiasoro2020_review, Gastiasoro2020, Ishizuka2018}.

Multipoles are classified by fundamental symmetries, namely spatial parity and time-reversal parity, into four classes: even-parity electric (EE), even-parity magnetic (EM), odd-parity electric (OE), and odd-parity magnetic (OM) multipoles.
In a pioneering work by Kozii and Fu~\cite{Kozii2015}, they proposed odd-parity superconductivity mediated by fluctuations of \textit{momentum-based} OE multipoles, which are represented by an electron's spin texture on the Fermi surface in spin-orbit-coupled systems.
However, superconductivity induced by the other classes of (EE, EM, and OM) multipole fluctuations, which has been investigated in specific models or materials~\cite{Koga2006, Ishizuka2018}, remains to be clarified in generic situations.
Furthermore, crystalline electric fields (CEFs) may cause higher anisotropy of superconductivity~\cite{Gastiasoro2020}, although Kozii and Fu considered only isotropic systems~\cite{Kozii2015}.

In this paper, we construct a general theory of \textit{ferroic} multipole-fluctuation-mediated superconductivity in spin-orbit-coupled systems.
First, a pairing interaction is formulated for all multipole fluctuations [Eq.~\eqref{eq:interaction_vertex_general}].
Using the formulation, we next calculate the induced pairing channels in both isotropic systems and crystalline systems.
In the isotropic case, the fluctuations of EE and OE multipoles yield attractive interactions not only in the $s$-wave channel but also in the anisotropic channel with the same symmetry as the multipoles.
However, no attractive interaction is induced by the EM and OM multipole fluctuations.
In the crystalline case, on the other hand, the CEF effect gives rise to significant effects on superconductivity.
For instance, anisotropic extended $s$-wave superconductivity may emerge owing to the magnetic multipole fluctuations as well as the electric ones.

This paper is constructed as follows.
First, in Sec.~\ref{sec:preparation}, we define momentum-based multipoles and introduce an interacting Hamiltonian induced by multipole fluctuations in the same manner as Ref.~\cite{Kozii2015}.
Next, we show a general formulation of the pairing interaction vertex for all (EE, EM, OE, and OM) multipoles in Sec.~\ref{sec:formulation}.
Then, we investigate multipole-fluctuation-induced pairings in isotropic systems (Sec.~\ref{sec:isotropic}) and crystalline systems (Sec.~\ref{sec:crystalline}).
Furthermore, in Sec.~\ref{sec:candidates}, we suggest unconventional superconductivity by multipole fluctuations in candidate materials: doped SrTiO$_3$, PrTi$_2$Al$_{20}$, Li$_2$(Pd,~Pt)$_3$B, and magnetic multipole systems.
Finally, a brief summary and discussion are given in Sec.~\ref{sec:summary}.

\section{Preparation}
\label{sec:preparation}
In this section, we introduce a momentum-based multipole operator and an effective Hamiltonian, which play an essential role in the paper.
Let $M$ be a magnetic point group symmetry in a disordered state, which is assumed to contain spatial inversion and time-reversal symmetry (TRS).
For simplicity, we restrict our discussion to a spin-orbit-coupled \textit{single-band} problem, where the band has a twofold (Kramers) degeneracy.

First, we define a multipole order parameter in order to discuss fluctuations.
A Hermitian operator $\hat{Q}$ is defined as
\begin{equation}
 \hat{Q} = \sum_{\bm{k}} \sum_{\alpha \beta} \Lambda_{\alpha \beta}(\bm{k}) c_{\bm{k} \alpha}^\dagger c_{\bm{k} \beta} \ \text{with} \ \Lambda^\dagger(\bm{k}) = \Lambda(\bm{k}),
 \label{eq:multipole_definition}
\end{equation}
where $\alpha$ and $\beta$ are pseudospin indices for the doubly degenerate states at every $\bm{k}$.
In the discussion, we choose the ``manifestly covariant Bloch basis'' used in Refs.~\cite{Fu2015, Kozii2015}.
Concretely speaking, the subscripts are exchanged under a time-reversal operation ($T$), while they are not changed under a spatial inversion ($I$):
\begin{align}
 T c_{\bm{k} \alpha}^\dagger T^{-1} &= \sum_{\beta} (i\sigma^y)_{\alpha \beta} c_{-\bm{k} \beta}^\dagger, \\
 I c_{\bm{k} \alpha}^\dagger I^{-1} &= c_{-\bm{k} \alpha}^\dagger,
\end{align}
where $\sigma^i$ is a Pauli matrix.
Based on the choice of the basis, the $2 \times 2$ matrix $\Lambda(\bm{k})$ has the following form,
\begin{equation}
 \Lambda(\bm{k}) = \begin{cases}
                   \psi_{\bm{k}} \sigma^0, \quad \psi_{\bm{k}} = \psi_{-\bm{k}} & \text{(EE)}, \\
                   \bm{c}_{\bm{k}} \cdot \bm{\sigma}, \quad \bm{c}_{\bm{k}} = \bm{c}_{-\bm{k}} & \text{(EM)}, \\
                   \bm{d}_{\bm{k}} \cdot \bm{\sigma}, \quad \bm{d}_{\bm{k}} = - \bm{d}_{-\bm{k}} & \text{(OE)}, \\
                   \phi_{\bm{k}} \sigma^0, \quad \phi_{\bm{k}} = - \phi_{-\bm{k}} & \text{(OM)},
                  \end{cases}
 \label{eq:multipole_form}
\end{equation}
where we take into account spatial parity and time-reversal parity of the multipole order.
$\psi_{\bm{k}}$, $\bm{c}_{\bm{k}}$, $\bm{d}_{\bm{k}}$, and $\phi_{\bm{k}}$ are real functions of $\bm{k}$ because of the Hermiticity of $\Lambda(\bm{k})$.
Equation~\eqref{eq:multipole_form} is consistent with the momentum representations of multipoles which are shown in the previous study~\cite{Watanabe2018, Hayami2018}.

Next, we introduce an effective $\hat{Q}$-$\hat{Q}$ interacting Hamiltonian,
\begin{equation}
 H_{\text{eff}} = \sum_{\bm{q}} V_{\bm{q}} \hat{Q}(\bm{q}) \hat{Q}(-\bm{q}),
 \label{eq:Q-Q_interaction}
\end{equation}
where $\hat{Q}(\bm{q}) = \hat{Q}^\dagger(-\bm{q})$ is the Fourier transform in the momentum ($\bm{q}$) space of the order parameter:
\begin{equation}
 \hat{Q}(\bm{q}) = \frac{1}{2} \sum_{\bm{k}} \sum_{\alpha \beta} \left\{ \Lambda_{\alpha \beta}(\bm{k} + \bm{q}) + \Lambda_{\alpha \beta}(\bm{k}) \right\} c_{\bm{k} + \bm{q} \alpha}^\dagger c_{\bm{k} \beta}.
 \label{eq:multipole_FT}
\end{equation}

Restricting the effective interaction [Eq.~\eqref{eq:Q-Q_interaction}] to pairing channels with zero center-of-mass momentum, we obtain the following reduced Hamiltonian,
\begin{equation}
 H_{\text{p}} = \sum_{\bm{k}, \bm{k}'} \sum_{\alpha \beta \gamma \delta} V_{\alpha \beta \delta \gamma}(\bm{k}, \bm{k}') c_{\bm{k} \alpha}^\dagger c_{-\bm{k} \beta}^\dagger c_{-\bm{k}' \gamma} c_{\bm{k}' \delta}, \label{eq:pairing_Hamiltonian}
\end{equation}
where the momentum- and pseudospin-dependent interaction vertex $V_{\alpha \beta \gamma \delta}(\bm{k}, \bm{k}')$ is given by
\begin{align}
 & V_{\alpha \beta \gamma \delta}(\bm{k}, \bm{k}') \notag \\
 &= \frac{1}{8} \bigl[ V_{\bm{k} - \bm{k}'} \{\Lambda(\bm{k}) + \Lambda(\bm{k}')\}_{\alpha \delta} \{\Lambda(-\bm{k}) + \Lambda(-\bm{k}')\}_{\beta \gamma} \notag \\
 & \qquad - V_{\bm{k} + \bm{k}'} \{\Lambda(\bm{k}) + \Lambda(-\bm{k}')\}_{\alpha \gamma} \{\Lambda(-\bm{k}) + \Lambda(\bm{k}')\}_{\beta \delta} \bigr]. \label{eq:interaction_vertex_exact}
\end{align}
The vertex represents an effective interaction between electrons induced by fluctuations of the multipole order.
As a simplest example, let us consider a charge (electric monopole) order $\Lambda(\bm{k}) = \sigma^0$.
Then, the vertex function is
\begin{align}
 V_{\alpha \beta \gamma \delta}(\bm{k}, \bm{k}') &= \frac{1}{4} (V_{\bm{k} - \bm{k}'} + V_{\bm{k} + \bm{k}'}) (i\sigma^y)_{\alpha \beta} (i\sigma^y)^\dagger_{\gamma \delta} \notag \\
 & \quad + \frac{1}{4} (V_{\bm{k} - \bm{k}'} - V_{\bm{k} + \bm{k}'}) (\bm{\sigma} i\sigma^y)_{\alpha \beta} \cdot (\bm{\sigma} i\sigma^y)^\dagger_{\gamma \delta}, \label{eq:interaction_vertex_isotropic_EM_exact}
\end{align}
where the first (second) term on the right-hand side (RHS) means spin-singlet (spin-triplet) pairing.
Furthermore, the interaction vertex for spin (magnetic dipole) fluctuations is similarly deduced as
\begin{align}
 V_{\alpha \beta \gamma \delta}(\bm{k}, \bm{k}') &= - \frac{3}{4} (V_{\bm{k} - \bm{k}'} + V_{\bm{k} + \bm{k}'}) (i\sigma^y)_{\alpha \beta} (i\sigma^y)^\dagger_{\gamma \delta} \notag \\
 & \quad + \frac{1}{4} (V_{\bm{k} - \bm{k}'} - V_{\bm{k} + \bm{k}'}) (\bm{\sigma} i\sigma^y)_{\alpha \beta} \cdot (\bm{\sigma} i\sigma^y)^\dagger_{\gamma \delta}, \label{eq:interaction_vertex_isotropic_MD_exact}
\end{align}
which is consistent with a well-known theory of spin-fluctuation-mediated superconductivity~\cite{Moriya2000, Yanase2003}.
This interaction results in, for example, spin-triplet superconductivity by ferromagnetic fluctuations~\cite{Nakajima1973, Fay1980, Hattori2014, Aoki2019_review}, and $d$-wave superconductivity by antiferromagnetic fluctuations~\cite{Miyake1986}.

Now we adopt a simple treatment used in Ref.~\cite{Kozii2015}, where $V_{\bm{k} \pm \bm{k}'} = V_0 \mp V_1 \hat{\bm{k}} \cdot \hat{\bm{k}'} + \dotsb$ is assumed to be approximated by the zeroth-order term $V_0 < 0$.
Note that the treatment is valid when the Fermi surface is relatively small, and $V_0$ should be negative so that \textit{ferroic} order of $\hat{Q}$ is favored in Eq.~\eqref{eq:Q-Q_interaction}.
Then, the interaction vertex [Eq.~\eqref{eq:interaction_vertex_exact}] is simplified as
\begin{align}
 & V_{\alpha \beta \gamma \delta}(\bm{k}, \bm{k}') \notag \\
 &= \frac{V_0}{8} \bigl[ \{\Lambda(\bm{k}) + \Lambda(\bm{k}')\}_{\alpha \delta} \{\Lambda(-\bm{k}) + \Lambda(-\bm{k}')\}_{\beta \gamma} \notag \\
 & \qquad - \{\Lambda(\bm{k}) + \Lambda(-\bm{k}')\}_{\alpha \gamma} \{\Lambda(-\bm{k}) + \Lambda(\bm{k}')\}_{\beta \delta} \bigr], \label{eq:interaction_vertex}
\end{align}
which is a key ingredient for determining the pairing symmetry in the superconducting state.
We here emphasize that the ``ordinary'' mechanism of unconventional superconductivity caused by spin fluctuations [Eq.~\eqref{eq:interaction_vertex_isotropic_MD_exact}] is qualitatively different from the mechanism described by Eq.~\eqref{eq:interaction_vertex}.
The momentum dependence of the effective spin-spin interaction $V_{\bm{q}}$ plays an essential role in the former theory, and it has been a canonical mechanism of unconventional superconductivity~\cite{Yanase2003}.
In the latter new mechanism, on the other hand, the momentum-based multipole $\Lambda(\bm{k})$ \textit{itself} has a momentum dependence, which is responsible for anisotropic pairing even when $V_{\bm{q}}$ is a constant $V_0$.
In the following sections, we do not take into account the momentum dependence of $V_{\bm{q}}$, but focus on the latter mechanism unless explicitly mentioned otherwise.
We calculate Eq.~\eqref{eq:interaction_vertex} and investigate what pairing symmetry is likely, in the vicinity of various multipole orders.

\section{General formulation}
\label{sec:formulation}
Now we derive a generic form of the vertex function $V_{\alpha \beta \gamma \delta}(\bm{k}, \bm{k}')$ [Eq.~\eqref{eq:interaction_vertex}] for all classes of multipole orders.
Consistent with the Landau theory of phase transitions, the multipole operator $\hat{Q}$ is classified by an irreducible representation (IR) of the symmetry in the disordered state.
Thus, Eq.~\eqref{eq:multipole_definition} is generalized for any IR ($\Gamma$) to
\begin{equation}
 \hat{Q}^{\Gamma n} = \sum_{\bm{k}} \sum_{\alpha \beta} \Lambda^{\Gamma n}_{\alpha \beta}(\bm{k}) c_{\bm{k} \alpha}^\dagger c_{\bm{k} \beta},
\end{equation}
where $n$ ($= 1, \dots, \dim\Gamma$) represents a basis index of the IR $\Gamma$.
In the vicinity of the $\Gamma$ multipole phase, the effective interaction [Eq.~\eqref{eq:Q-Q_interaction}] is naturally extended to
\begin{equation}
 H_{\text{eff}}^{\Gamma} = \sum_{n = 1}^{\dim\Gamma} \sum_{\bm{q}} V_{\bm{q}} \hat{Q}^{\Gamma n}(\bm{q}) \hat{Q}^{\Gamma n}(-\bm{q}).
\end{equation}
Then, the interaction vertex under the approximation $V_{\bm{k} \pm \bm{k}'} \simeq V_0$ is formulated as
\begin{widetext}
 \begin{equation}
  V_{\alpha \beta \gamma \delta}(\bm{k}, \bm{k}') =
  \begin{dcases}
   - \frac{|V_0|}{8} \sum_{n} \{(\psi^{\Gamma n}_{\bm{k}})^2 + (\psi^{\Gamma n}_{\bm{k}'})^2\} (i\sigma^y)_{\alpha \beta} (i\sigma^y)^\dagger_{\gamma \delta} - \frac{|V_0|}{4} \sum_{n} (\psi^{\Gamma n}_{\bm{k}} i\sigma^y)_{\alpha \beta} (\psi^{\Gamma n}_{\bm{k}'} i\sigma^y)^\dagger_{\gamma \delta} & \text{(EE),} \\
   \frac{|V_0|}{8} \sum_{n} \{|\bm{c}^{\Gamma n}_{\bm{k}}|^2 + |\bm{c}^{\Gamma n}_{\bm{k}'}|^2\} (i\sigma^y)_{\alpha \beta} (i\sigma^y)^\dagger_{\gamma \delta} + \frac{|V_0|}{4} \sum_{n} (\bm{c}^{\Gamma n}_{\bm{k}} i\sigma^y)_{\alpha \beta} \cdot (\bm{c}^{\Gamma n}_{\bm{k}'} i\sigma^y)^\dagger_{\gamma \delta} & \text{(EM),} \\
   \begin{aligned}[b]
    & - \frac{|V_0|}{8} \sum_{n} \{|\bm{d}^{\Gamma n}_{\bm{k}}|^2 + |\bm{d}^{\Gamma n}_{\bm{k}'}|^2\} (i\sigma^y)_{\alpha \beta} (i\sigma^y)^\dagger_{\gamma \delta} \\
    & \quad - \frac{|V_0|}{4} \sum_{n} (\bm{d}^{\Gamma n}_{\bm{k}} \cdot \bm{\sigma} i\sigma^y)_{\alpha \beta} (\bm{d}^{\Gamma n}_{\bm{k}'} \cdot \bm{\sigma} i\sigma^y)^\dagger_{\gamma \delta} + \frac{|V_0|}{4} \sum_{n} (\bm{d}^{\Gamma n}_{\bm{k}} \times \bm{\sigma} i\sigma^y)_{\alpha \beta} \cdot (\bm{d}^{\Gamma n}_{\bm{k}'} \times \bm{\sigma} i\sigma^y)^\dagger_{\gamma \delta}
   \end{aligned} & \text{(OE),} \\
   \frac{|V_0|}{8} \sum_{n} \{(\phi^{\Gamma n}_{\bm{k}})^2 + (\phi^{\Gamma n}_{\bm{k}'})^2\} (i\sigma^y)_{\alpha \beta} (i\sigma^y)^\dagger_{\gamma \delta} + \frac{|V_0|}{4} \sum_{n} (\phi^{\Gamma n}_{\bm{k}} \bm{\sigma} i\sigma^y)_{\alpha \beta} \cdot (\phi^{\Gamma n}_{\bm{k}'} \bm{\sigma} i\sigma^y)^\dagger_{\gamma \delta} & \text{(OM),}
  \end{dcases}
  \label{eq:interaction_vertex_general}
 \end{equation}
\end{widetext}
where we use $V_0 < 0$.
Equation~\eqref{eq:interaction_vertex_general} is one of the main results in this paper.
Pairing interactions induced by EE, EM, OE, and OM multipole fluctuations are shown in this order.
Using this, we can discuss the possible superconducting instability in various circumstances.
The isotropic systems are studied in Sec.~\ref{sec:isotropic}, while the crystalline systems are investigated in Sec.~\ref{sec:crystalline}.

\section{Interaction vertex in isotropic systems}
\label{sec:isotropic}
In this section, we clarify the stable pairing states in \textit{isotropic} systems, where the magnetic ``point group'' is represented by the rotation group: $M = O(3) + O(3) T$.
In the case, the IR and its bases are given by the total angular momentum: $\Gamma = J$ ($= 0, 1, 2, \dotsc$) and $n = M$ ($= - J, - J + 1 \dots, J$).
Using the bases, we first review the previously suggested spin-triplet superconductivity induced by an OE multipole fluctuation~\cite{Kozii2015}.
Furthermore, the theory is extended for the other classes (OM, EE, and EM) of multipole fluctuations.

\subsection{Review: OE multipoles}
First, we revisit the OE-fluctuation-mediated superconductivity proposed in the previous study by Kozii and Fu~\cite{Kozii2015}.
We here set a basis function of the OE multipole $\Lambda_{\text{OE}}^{J M}(\bm{k}) = \bm{d}^{J M}_{\bm{k}} \cdot \bm{\sigma}$ for each angular momentum.
In Refs.~\cite{Fu2015, Kozii2015}, three total angular momenta $J = 0$, $1$, and $2$ are considered:
\begin{subequations}
 \label{eq:OE_multipole_isotropic}
 \begin{align}
  \Lambda_{\text{OE}}^{0 0}(\bm{k}) &= \hat{\bm{k}} \cdot \bm{\sigma}, \label{eq:OE_multipole_isotropic_J0} \\
  \Lambda_{\text{OE}}^{1 i}(\bm{k}) &= (\hat{\bm{k}} \times \bm{\sigma})^i, \label{eq:OE_multipole_isotropic_J1} \\
  \Lambda_{\text{OE}}^{2, i j}(\bm{k}) &= \hat{k}^i \sigma^j + \hat{k}^j \sigma^i - \frac{2}{3}(\hat{\bm{k}} \cdot \bm{\sigma}) \delta^{i j},
 \end{align}
\end{subequations}
where the $L = 1$ orbital angular momentum, namely the linear $\bm{k}$ dependence, is assumed.%
\footnote{For $J = 1$ and $2$, classification of the OE operators by $M$ is given by
\begin{align*}
 \Lambda_{\text{OE}}^{1 M}(\bm{k}) &\sim
 \begin{cases}
  (\hat{\bm{k}} \times \bm{\sigma})^z & M = 0, \\
  \mp \frac{1}{\sqrt{2}} \{(\hat{\bm{k}} \times \bm{\sigma})^x \pm i (\hat{\bm{k}} \times \bm{\sigma})^y\} & M = \pm 1,
 \end{cases} \displaybreak[2] \\
 \Lambda_{\text{OE}}^{2 M}(\bm{k}) &\sim
 \begin{cases}
  \sqrt{\frac{2}{3}} (2 \hat{k}^z \sigma^z - \hat{k}^x \sigma^x - \hat{k}^y \sigma^y) & M = 0, \\
  \mp \left\{ (\hat{k}^z \sigma^x + \hat{k}^x \sigma^z) \pm i (\hat{k}^y \sigma^z + \hat{k}^z \sigma^y) \right\} & M = \pm 1, \\
  (\hat{k}^x \sigma^x - \hat{k}^y \sigma^y) \pm i (\hat{k}^x \sigma^y + \hat{k}^y \sigma^x) & M = \pm 2.
 \end{cases}
\end{align*}
Although some of these functions are non-Hermitian, the Hermitian forms in Eq.~\eqref{eq:OE_multipole_isotropic} can be straightforwardly obtained by appropriately transforming the bases.}
The momentum dependence of multipole operators causes unconventional superconducting instability even when $V_{\bm{k} \pm \bm{k}'}$ is approximated by the zeroth order $V_0$.
For example, we suppose the vicinity of the $J = 1$ order, which is spontaneous emergence of a Rashba structure in the momentum space [Eq.~\eqref{eq:OE_multipole_isotropic_J1}].
Then, the vertex function [Eq.~\eqref{eq:interaction_vertex_general}(OE)] is calculated as
\begin{align}
 V_{\alpha \beta \gamma \delta}(\bm{k}, \bm{k}') &= - \frac{|V_0|}{2} (i\sigma^y)_{\alpha \beta} (i\sigma^y)^\dagger_{\gamma \delta} \notag \\
 & \quad + \frac{|V_0|}{3} \{\Lambda_{\text{OE}}^{0 0}(\bm{k}) i\sigma^y\}_{\alpha \beta} \{\Lambda_{\text{OE}}^{0 0}(\bm{k}') i\sigma^y\}^\dagger_{\gamma \delta} \notag \\
 & \quad - \frac{|V_0|}{8} \sum_{i} \{\Lambda_{\text{OE}}^{1 i}(\bm{k}) i\sigma^y\}_{\alpha \beta} \{\Lambda_{\text{OE}}^{1 i}(\bm{k}') i\sigma^y\}^\dagger_{\gamma \delta} \notag \\
 & \quad + \frac{|V_0|}{16} \sum_{i, j} \{\Lambda_{\text{OE}}^{2, i j}(\bm{k}) i\sigma^y\}_{\alpha \beta} \{\Lambda_{\text{OE}}^{2, i j}(\bm{k}') i\sigma^y\}^\dagger_{\gamma \delta},
 \label{eq:interaction_vertex_isotropic_ED}
\end{align}
where the third term for spin-triplet pairing as well as the first term for spin-singlet ($s$-wave) pairing give an attractive interaction, while the second and fourth spin-triplet terms are repulsive.
Because in realistic superconductors, the screened Coulomb repulsion should make a pair-breaking effect in the $s$-wave channel~\cite{Fu2010, Brydon2014, Kozii2015, Kozii2019}, the odd-parity pairing due to the third term may be energetically favored by the fluctuation of OE multipoles.
Similar results are obtained for the other $J = 0$ and $2$ multipoles~\cite{Kozii2015}; in general, the induced odd-parity pairing possesses the same symmetry as the fluctuating OE multipole, which is easily seen in the second term of Eq.~\eqref{eq:interaction_vertex_general}(OE).

\subsection{Result: OM multipoles}
Next, we investigate OM multipoles.
We consider a basis function of the OM order for each angular momentum, in a similar way to OE multipoles.
Restricting the discussion to the $L = 1$ orbital, the basis function $\Lambda_{\text{OM}}^{J M}(\bm{k}) = \phi^{J M}_{\bm{k}} \sigma^0$ has components only for $J = 1$ total angular momentum.
Indeed, the $J = 1$ basis functions are given by
\begin{equation}
 \Lambda_{\text{OM}}^{1 i}(\bm{k}) = \hat{k}^i \sigma^0.
\end{equation}
Therefore, we only calculate Eq.~\eqref{eq:interaction_vertex_general}(OM) for the $J = 1$ bases:
\begin{align}
 V_{\alpha \beta \gamma \delta}(\bm{k}, \bm{k}') &= \frac{|V_0|}{4} (i\sigma^y)_{\alpha \beta} (i\sigma^y)^\dagger_{\gamma \delta} \notag \\
 & \quad + \frac{|V_0|}{12} \{\Lambda_{\text{OE}}^{0 0}(\bm{k}) i\sigma^y\}_{\alpha \beta} \{\Lambda_{\text{OE}}^{0 0}(\bm{k}') i\sigma^y\}^\dagger_{\gamma \delta} \notag \\
 & \quad + \frac{|V_0|}{8} \sum_{i} \{\Lambda_{\text{OE}}^{1 i}(\bm{k}) i\sigma^y\}_{\alpha \beta} \cdot \{\Lambda_{\text{OE}}^{1 i}(\bm{k}') i\sigma^y\}^\dagger_{\gamma \delta} \notag \\
 & \quad + \frac{|V_0|}{16} \sum_{i, j} \{\Lambda_{\text{OE}}^{2, i j}(\bm{k}) i\sigma^y \}_{\alpha \beta} \{\Lambda_{\text{OE}}^{2, i j}(\bm{k}') i\sigma^y \}^\dagger_{\gamma \delta}.
 \label{eq:interaction_vertex_isotropic_MTD}
\end{align}
As apparent from the above equation, fluctuations of OM multipoles in the isotropic space unfortunately cause the \textit{repulsive} pairing interaction irrespective of the spatial parity (spin-singlet or spin-triplet) of superconductivity.
In this case, multipole-fluctuation-mediated superconductivity is unlikely.
Later, we show that an anisotropic superconductivity may be stabilized in crystalline systems.

\subsection{Result: EE multipole}
Now let us move on to even-parity multipole orders.
The simplest (lowest-order) basis function of EE multipole order is an electric monopole (charge) with $J = 0$: $\Lambda_{\text{EE}}^{00}(\bm{k}) = \sigma^0$.
Then, the interaction vertex [Eq.~\eqref{eq:interaction_vertex_general}(EE)] due to the charge fluctuation is
\begin{equation}
 V_{\alpha \beta \gamma \delta}(\bm{k}, \bm{k}') = - \frac{|V_0|}{2} (i\sigma^y)_{\alpha \beta} (i\sigma^y)^\dagger_{\gamma \delta},
 \label{eq:interaction_vertex_isotropic_EM}
\end{equation}
which represents an attractive interaction for the isotropic $s$-wave pairing.
Thus, the charge fluctuation induces conventional $s$-wave superconductivity, and this is true beyond the condition $V_{\bm{k} \pm \bm{k}'} \simeq V_0$ [Eq.~\eqref{eq:interaction_vertex_isotropic_EM_exact}].

On the other hand, the result is changed in the vicinity of a higher-order EE multipole state.
Considering the second-lowest-order EE multipole, namely electric quadrupoles with $J = 2$, the basis functions are represented as $\Lambda_{\text{EE}}^{2 i}(\bm{k}) = \psi^{2 i}_{\bm{k}} \sigma^0$ with
\begin{align}
 \{\psi^{2 i}_{\bm{k}}\} = \biggl\{ & \frac{1}{2} \left(2 (\hat{k}^z)^2 - (\hat{k}^x)^2 - (\hat{k}^y)^2\right), \notag \\
 & \sqrt{3} \hat{k}^z \hat{k}^x, \, \sqrt{3} \hat{k}^y \hat{k}^z, \notag \\
 & \frac{\sqrt{3}}{2} \left((\hat{k}^x)^2 - (\hat{k}^y)^2\right), \, \sqrt{3} \hat{k}^x \hat{k}^y \biggr\}.
\end{align}
Substituting the fivefold-degenerate representation into Eq.~\eqref{eq:interaction_vertex_general}(EE), the interaction function is given by
\begin{align}
 V_{\alpha \beta \gamma \delta}(\bm{k}, \bm{k}') &= - \frac{|V_0|}{4} (i\sigma^y)_{\alpha \beta} (i\sigma^y)^\dagger_{\gamma \delta} \notag \\
 & \quad - \frac{|V_0|}{4} \sum_{i} \{\Lambda_{\text{EE}}^{2 i}(\bm{k}) i\sigma^y\}_{\alpha \beta} \{\Lambda_{\text{EE}}^{2 i}(\bm{k}') i\sigma^y\}^\dagger_{\gamma \delta},
 \label{eq:interaction_vertex_isotropic_EQ}
\end{align}
where the second term means an attractive $d$-wave pairing, while the first term favors $s$-wave superconductivity.
Generally speaking, a fluctuation of higher-order EE multipoles induces an anisotropic pairing with the same symmetry as the multipoles [the second term in Eq.~\eqref{eq:interaction_vertex_general}(EE)], as well as a conventional $s$-wave pairing [the first term in Eq.~\eqref{eq:interaction_vertex_general}(EE)].

\subsection{Result: EM multipole}
Finally we consider EM multipole fluctuations.
The lowest-order basis functions are magnetic dipole (spin) with $J = 1$: $\Lambda_{\text{EM}}^{1 i}(\bm{k}) = \sigma^i$ ($i = x, y, z$).
Therefore, the interaction vertex [Eq.~\eqref{eq:interaction_vertex_general}(EM)] is given by
\begin{equation}
 V_{\alpha \beta \gamma \delta}(\bm{k}, \bm{k}') = \frac{3 |V_0|}{2} (i\sigma^y)_{\alpha \beta} (i\sigma^y)^\dagger_{\gamma \delta},
 \label{eq:interaction_vertex_isotropic_MD}
\end{equation}
which means that any superconducting phase is not stabilized.
This is an artifact of our assumption $V_{\bm{k} \pm \bm{k}'} \simeq V_0$.
Beyond the condition, the vertex function has a well-known form of spin-fluctuation-mediated interaction [Eq.~\eqref{eq:interaction_vertex_isotropic_MD_exact}], where the momentum dependence of $V_{\bm{q}}$ plays an important role for stabilizing superconductivity~\cite{Yanase2003}.
As we mentioned before, however, we do not touch this mechanism and focus on another mechanism due to momentum dependence of multipole operators.

Within $V_{\bm{k} \pm \bm{k}'} \simeq V_0$, higher-order EM multipoles also mediate no attractive pairing.
Indeed, fluctuation of magnetic octupole ($J = 3$) orders $\Lambda_{\text{EM}}^{3 i}(\bm{k}) = \bm{c}^{3 i}_{\bm{k}} \cdot \bm{\sigma}$ with~\cite{Watanabe2018, Hayami2018},
\begin{align}
 \{\bm{c}^{3 i}_{\bm{k}}\} =
 \biggl\{ & \frac{3}{2} \left( - 2 \hat{k}^x \hat{k}^z, \, - 2 \hat{k}^y \hat{k}^z, \, 2 (\hat{k}^z)^2 - (\hat{k}^x)^2 - (\hat{k}^y)^2 \right), \notag \\
 & - \frac{\sqrt{6}}{4} \left( 3 (\hat{k}^x)^2 + (\hat{k}^y)^2 - 4 (\hat{k}^z)^2, \, 2 \hat{k}^x \hat{k}^y, \, - 8 \hat{k}^x \hat{k}^z \right), \notag \\
 & - \frac{\sqrt{6}}{4} \left( 2 \hat{k}^x \hat{k}^y, \, (\hat{k}^x)^2 + 3 (\hat{k}^y)^2 - 4 (\hat{k}^z)^2, \, - 8 \hat{k}^y \hat{k}^z \right), \notag \\
 & \frac{\sqrt{15}}{2} \left( 2 \hat{k}^x \hat{k}^z, \, - 2 \hat{k}^y \hat{k}^z, \, (\hat{k}^x)^2 - (\hat{k}^y)^2 \right), \notag \\
 & \frac{\sqrt{15}}{2} \left( 2 \hat{k}^y \hat{k}^z, \, 2 \hat{k}^x \hat{k}^z, \, 2 \hat{k}^x \hat{k}^y \right), \notag \\
 & \frac{3\sqrt{10}}{4} \left( (\hat{k}^x)^2 - (\hat{k}^y)^2, \, - 2 \hat{k}^x \hat{k}^y, \, 0 \right), \notag \\
 & \frac{3\sqrt{10}}{4} \left( 2 \hat{k}^x \hat{k}^y, \, (\hat{k}^x)^2 - (\hat{k}^y)^2, \, 0 \right) \biggr\},
\end{align}
gives rise to the following interaction vertex:
\begin{align}
 V_{\alpha \beta \gamma \delta}(\bm{k}, \bm{k}') &= \frac{21 |V_0|}{4} (i\sigma^y)_{\alpha \beta} (i\sigma^y)^\dagger_{\gamma \delta} \notag \\
 & \quad + \frac{21 |V_0|}{4} \sum_{i} \{\Lambda_{\text{EE}}^{2 i}(\bm{k}) i\sigma^y\}_{\alpha \beta} \{\Lambda_{\text{EE}}^{2 i}(\bm{k}') i\sigma^y\}^\dagger_{\gamma \delta}.
 \label{eq:interaction_vertex_isotropic_MO}
\end{align}
Although Eq.~\eqref{eq:interaction_vertex_isotropic_MO} has a form similar to that of Eq.~\eqref{eq:interaction_vertex_isotropic_EQ}, all the terms have positive coefficients, revealing the absence of attractive pairing.

\section{Interaction vertex under crystal symmetry}
\label{sec:crystalline}
In the previous section, we have elucidated superconductivity mediated by multipole fluctuations in \textit{isotropic} systems where multipoles are classified by total angular momentum $J$.
In real superconductors, on the other hand, CEF causes splitting of degeneracy in the $J$ manifold and induces further anisotropic interaction.

In the following subsections, we calculate the interaction vertex \textit{under CEFs}.
Systems with three high-symmetry crystal point groups $D_{4h}$, $D_{6h}$, and $O_h$ are analyzed in a comprehensive manner, following the previous classification theories of unconventional superconductivity~\cite{Sigrist-Ueda} and multipole order~\cite{Watanabe2018}.
We show universal relations between the symmetry of multipole order and superconductivity.
They are also expected to be valid in systems of lower crystalline symmetry.

\subsection{Review: classification of multipole order}
\label{sec:crystalline-multipole}
First of all, we revisit the classification theory of multipole order parameters under the CEF~\cite{Watanabe2018, Hayami2018}.
From the viewpoint of representation theory, the CEF causes two effects on multipole orders: (i) degeneracy splitting among \textit{same}-order multipoles and (ii) representation merging among \textit{different}-order multipoles.
For the help of understanding, let us consider OE dipole moments $\{x, y, z\}$, which are basis functions of the threefold-degenerate $J = 1$ representation in isotropic systems, as an example.
When the CEF with tetragonal $D_{4h}$ symmetry is switched on, (i) the threefold degeneracy splits into two IRs of $D_{4h}$, namely, a nondegenerate $A_{2u}$ with the basis $\{z\}$ and a doubly degenerate $E_u$ with $\{x, y\}$.
Furthermore, since $D_{4h}$ is a finite group, (ii) these $J = 1$ multipoles and higher-order ($J = 3, 5, \dotsc$) multipoles are merged into the same IR; for example, the above $A_{2u}$ IR has basis functions of an electric octupole $Q_{3 0} \sim \frac{1}{2} (5z^2 - 3r^2) z$ as well as the electric dipole $Q_{1 0} \sim z$.

For the above reasons, classification of multipoles in crystalline systems is significantly different from that in isotropic systems.
Indeed, recent studies have shown the list of IRs in crystal point groups and the corresponding basis functions of multipole moments~\cite{Watanabe2018, Hayami2018}.
Parts of their classification tables are reprinted in Tables~\ref{tab:multipole_basis_D4h}--\ref{tab:multipole_basis_Oh} of Appendix~\ref{ap:classification_multipole}, which provide lowest-order basis functions in the real-space coordinates $\bm{r} = (x, y, z)$ and those in the momentum-space coordinates $\hat{\bm{k}} = (\hat{k}^x, \hat{k}^y, \hat{k}^z)$.
We denote IRs with even (odd) time-reversal parity by $\Gamma^+$ ($\Gamma^-$); in other words, electric (magnetic) multipole moments are represented by $\Gamma^+$ ($\Gamma^-$).

Tables~\ref{tab:multipole_basis_D4h}--\ref{tab:multipole_basis_Oh} are useful to identify correspondence between the fluctuating real-space multipoles and the momentum-space basis functions used in our theory [Eq.~\eqref{eq:interaction_vertex_general}].
For example, when we discuss a ferroelectric fluctuation along the $z$ axis in tetragonal superconductors, we see Table~\ref{tab:multipole_basis_D4h}(OE), which shows that the electric dipole ($z$) belongs to the $A_{2u}^+$ IR, and the corresponding momentum-space basis is $\bm{d}^{A_{2u}}_{\bm{k}} = \hat{k}^x \hat{\bm{y}} - \hat{k}^y \hat{\bm{x}}$.

\subsection{Result: classification of interaction vertex under CEF}
Using basis functions in Tables~\ref{tab:multipole_basis_D4h}--\ref{tab:multipole_basis_Oh}, we calculate the interaction vertex in Eq.~\eqref{eq:interaction_vertex_general} and decompose it into IRs of the corresponding point group.
We accomplish the calculation for all IRs in the point groups $D_{4h}$, $D_{6h}$, and $O_h$, with the help of GTPack~\cite{gtpack1, gtpack2}, a free Mathematica group theory package.
The irreducible decomposition of the vertex deduced from the multipole fluctuations is given in Tables~\ref{tab:vertex_D4h}--\ref{tab:vertex_Oh}, for all even-parity/odd-parity electric/magnetic basis functions.

\begin{table*}[tbp]
 \centering
 \caption{Irreducible decomposition of interaction vertex induced by multipole fluctuations for all IRs under $D_{4h}$ CEF. The IR of the multipole is denoted by $\Gamma$. The middle column shows the interaction vertex for the four classes (EE, EM, OE, and OM) of multipoles. We obtain the same form within the symmetry class specified by spatial parity and time-reversal parity. An explicit form for each IR is given by assigning the basis functions $\psi_{\bm{k}}$ (EE), $\bm{c}_{\bm{k}}$ (EM), $\bm{d}_{\bm{k}}$ (OE), and $\phi_{\bm{k}}$ (OM) listed in Tables~\ref{tab:multipole_basis_D4h}--\ref{tab:multipole_basis_Oh}. The right column represents the IRs of pairing channels.}
 \label{tab:vertex_D4h}
 \begin{tabularx}{\linewidth}{cXlXl} \hline\hline
  Multipole ($\Gamma$) && $V_{\alpha \beta \gamma \delta}(\bm{k}, \bm{k}') / |V_0|$ && Decomposition \\ \hline
  EE ($\Gamma_g^+$) && $- \frac{1}{8} \sum_{n} \{(\psi^{\Gamma n}_{\bm{k}})^2 + (\psi^{\Gamma n}_{\bm{k}'})^2\} (i\sigma^y)_{\alpha \beta} (i\sigma^y)^\dagger_{\gamma \delta}$ && $A_{1g}$ \\
  && $- \frac{1}{4} \sum_{n} (\psi^{\Gamma n}_{\bm{k}} i\sigma^y)_{\alpha \beta} (\psi^{\Gamma n}_{\bm{k}'} i\sigma^y)^\dagger_{\gamma \delta}$ && $\Gamma_g$ \\
  \\
  EM ($\Gamma_g^-$) && $+ \frac{1}{8} \sum_{n} \{|\bm{c}^{\Gamma n}_{\bm{k}}|^2 + |\bm{c}^{\Gamma n}_{\bm{k}'}|^2\} (i\sigma^y)_{\alpha \beta} (i\sigma^y)^\dagger_{\gamma \delta}$ && $A_{1g}$ \\
  && $+ \frac{1}{4} \sum_{n} \{(\bm{c}^{\Gamma n}_{\bm{k}})^z i\sigma^y\}_{\alpha \beta} \{(\bm{c}^{\Gamma n}_{\bm{k}'})^z i\sigma^y\}^\dagger_{\gamma \delta}$ && $\Gamma_g \times A_{2g}$ \\
  && $+ \frac{1}{4} \sum_{n} \sum_{j = x, y} \{(\bm{c}^{\Gamma n}_{\bm{k}})^j i\sigma^y\}_{\alpha \beta} \{(\bm{c}^{\Gamma n}_{\bm{k}'})^j i\sigma^y\}^\dagger_{\gamma \delta}$ && $\Gamma_g \times E_g$ \\
  \\
  OE ($\Gamma_u^+$) && $- \frac{1}{8} \sum_{n} \{|\bm{d}^{\Gamma n}_{\bm{k}}|^2 + |\bm{d}^{\Gamma n}_{\bm{k}'}|^2\} (i\sigma^y)_{\alpha \beta} (i\sigma^y)^\dagger_{\gamma \delta}$ && $A_{1g}$ \\
  && $- \frac{1}{4} \sum_{n} (\bm{d}^{\Gamma n}_{\bm{k}} \cdot \bm{\sigma} i\sigma^y)_{\alpha \beta} (\bm{d}^{\Gamma n}_{\bm{k}'} \cdot \bm{\sigma} i\sigma^y)^\dagger_{\gamma \delta}$ && $\Gamma_u$ \\
  && $+ \frac{1}{4} \sum_{n} \{(\bm{d}^{\Gamma n}_{\bm{k}} \times \bm{\sigma})^z i\sigma^y\}_{\alpha \beta} \{(\bm{d}^{\Gamma n}_{\bm{k}'} \times \bm{\sigma})^z i\sigma^y\}^\dagger_{\gamma \delta}$ && $\Gamma_u \times A_{2g}$ \\
  && $+ \frac{1}{4} \sum_{n} \sum_{j = x, y} \{(\bm{d}^{\Gamma n}_{\bm{k}} \times \bm{\sigma})^j i\sigma^y\}_{\alpha \beta} \{(\bm{d}^{\Gamma n}_{\bm{k}'} \times \bm{\sigma})^j i\sigma^y\}^\dagger_{\gamma \delta}$ && $\Gamma_u \times E_g$ \\
  \\
  OM ($\Gamma_u^-$) && $+ \frac{1}{8} \sum_{n} \{(\phi^{\Gamma n}_{\bm{k}})^2 + (\phi^{\Gamma n}_{\bm{k}'})^2\} (i\sigma^y)_{\alpha \beta} (i\sigma^y)^\dagger_{\gamma \delta}$ && $A_{1g}$ \\
  && $+ \frac{1}{4} \sum_{n} (\phi^{\Gamma n}_{\bm{k}} \sigma^z i\sigma^y)_{\alpha \beta} (\phi^{\Gamma n}_{\bm{k}'} \sigma^z i\sigma^y)^\dagger_{\gamma \delta}$ && $\Gamma_u \times A_{2g}$ \\
  && $+ \frac{1}{4} \sum_{n} \sum_{j = x, y} (\phi^{\Gamma n}_{\bm{k}} \sigma^j i\sigma^y)_{\alpha \beta} (\phi^{\Gamma n}_{\bm{k}'} \sigma^j i\sigma^y)^\dagger_{\gamma \delta}$ && $\Gamma_u \times E_g$ \\ \hline\hline
 \end{tabularx}
\end{table*}

\begin{table*}[tbp]
 \centering
 \caption{Irreducible decomposition of interaction vertex induced by multipole fluctuations for all IRs under $D_{6h}$ CEF.}
 \label{tab:vertex_D6h}
 \begin{tabularx}{\linewidth}{cXlXl} \hline\hline
  Multipole ($\Gamma$) && $V_{\alpha \beta \gamma \delta}(\bm{k}, \bm{k}') / |V_0|$ && Decomposition \\ \hline
  EE ($\Gamma_g^+$) && $- \frac{1}{8} \sum_{n} \{(\psi^{\Gamma n}_{\bm{k}})^2 + (\psi^{\Gamma n}_{\bm{k}'})^2\} (i\sigma^y)_{\alpha \beta} (i\sigma^y)^\dagger_{\gamma \delta}$ && $A_{1g}$ \\
  && $- \frac{1}{4} \sum_{n} (\psi^{\Gamma n}_{\bm{k}} i\sigma^y)_{\alpha \beta} (\psi^{\Gamma n}_{\bm{k}'} i\sigma^y)^\dagger_{\gamma \delta}$ && $\Gamma_g$ \\
  \\
  EM ($\Gamma_g^-$) && $+ \frac{1}{8} \sum_{n} \{|\bm{c}^{\Gamma n}_{\bm{k}}|^2 + |\bm{c}^{\Gamma n}_{\bm{k}'}|^2\} (i\sigma^y)_{\alpha \beta} (i\sigma^y)^\dagger_{\gamma \delta}$ && $A_{1g}$ \\
  && $+ \frac{1}{4} \sum_{n} \{(\bm{c}^{\Gamma n}_{\bm{k}})^z i\sigma^y\}_{\alpha \beta} \{(\bm{c}^{\Gamma n}_{\bm{k}'})^z i\sigma^y\}^\dagger_{\gamma \delta}$ && $\Gamma_g \times A_{2g}$ \\
  && $+ \frac{1}{4} \sum_{n} \sum_{j = x, y} \{(\bm{c}^{\Gamma n}_{\bm{k}})^j i\sigma^y\}_{\alpha \beta} \{(\bm{c}^{\Gamma n}_{\bm{k}'})^j i\sigma^y\}^\dagger_{\gamma \delta}$ && $\Gamma_g \times E_{1g}$ \\
  \\
  OE ($\Gamma_u^+$) && $- \frac{1}{8} \sum_{n} \{|\bm{d}^{\Gamma n}_{\bm{k}}|^2 + |\bm{d}^{\Gamma n}_{\bm{k}'}|^2\} (i\sigma^y)_{\alpha \beta} (i\sigma^y)^\dagger_{\gamma \delta}$ && $A_{1g}$ \\
  && $- \frac{1}{4} \sum_{n} (\bm{d}^{\Gamma n}_{\bm{k}} \cdot \bm{\sigma} i\sigma^y)_{\alpha \beta} (\bm{d}^{\Gamma n}_{\bm{k}'} \cdot \bm{\sigma} i\sigma^y)^\dagger_{\gamma \delta}$ && $\Gamma_u$ \\
  && $+ \frac{1}{4} \sum_{n} \{(\bm{d}^{\Gamma n}_{\bm{k}} \times \bm{\sigma})^z i\sigma^y\}_{\alpha \beta} \{(\bm{d}^{\Gamma n}_{\bm{k}'} \times \bm{\sigma})^z i\sigma^y\}^\dagger_{\gamma \delta}$ && $\Gamma_u \times A_{2g}$ \\
  && $+ \frac{1}{4} \sum_{n} \sum_{j = x, y} \{(\bm{d}^{\Gamma n}_{\bm{k}} \times \bm{\sigma})^j i\sigma^y\}_{\alpha \beta} \{(\bm{d}^{\Gamma n}_{\bm{k}'} \times \bm{\sigma})^j i\sigma^y\}^\dagger_{\gamma \delta}$ && $\Gamma_u \times E_{1g}$ \\
  \\
  OM ($\Gamma_u^-$) && $+ \frac{1}{8} \sum_{n} \{(\phi^{\Gamma n}_{\bm{k}})^2 + (\phi^{\Gamma n}_{\bm{k}'})^2\} (i\sigma^y)_{\alpha \beta} (i\sigma^y)^\dagger_{\gamma \delta}$ && $A_{1g}$ \\
  && $+ \frac{1}{4} \sum_{n} (\phi^{\Gamma n}_{\bm{k}} \sigma^z i\sigma^y)_{\alpha \beta} (\phi^{\Gamma n}_{\bm{k}'} \sigma^z i\sigma^y)^\dagger_{\gamma \delta}$ && $\Gamma_u \times A_{2g}$ \\
  && $+ \frac{1}{4} \sum_{n} \sum_{j = x, y} (\phi^{\Gamma n}_{\bm{k}} \sigma^j i\sigma^y)_{\alpha \beta} (\phi^{\Gamma n}_{\bm{k}'} \sigma^j i\sigma^y)^\dagger_{\gamma \delta}$ && $\Gamma_u \times E_{1g}$ \\ \hline\hline
 \end{tabularx}
\end{table*}

\begin{table*}[tbp]
 \centering
 \caption{Irreducible decomposition of interaction vertex induced by multipole fluctuations for all IRs under $O_h$ CEF.}
 \label{tab:vertex_Oh}
 \begin{tabularx}{\linewidth}{cXlXl} \hline\hline
  Multipole ($\Gamma$) && $V_{\alpha \beta \gamma \delta}(\bm{k}, \bm{k}') / |V_0|$ && Decomposition \\ \hline
  EE ($\Gamma_g^+$) && $- \frac{1}{8} \sum_{n} \{(\psi^{\Gamma n}_{\bm{k}})^2 + (\psi^{\Gamma n}_{\bm{k}'})^2\} (i\sigma^y)_{\alpha \beta} (i\sigma^y)^\dagger_{\gamma \delta}$ && $A_{1g}$ \\
  && $- \frac{1}{4} \sum_{n} (\psi^{\Gamma n}_{\bm{k}} i\sigma^y)_{\alpha \beta} (\psi^{\Gamma n}_{\bm{k}'} i\sigma^y)^\dagger_{\gamma \delta}$ && $\Gamma_g$ \\
  \\
  EM ($\Gamma_g^-$) && $+ \frac{1}{8} \sum_{n} \{|\bm{c}^{\Gamma n}_{\bm{k}}|^2 + |\bm{c}^{\Gamma n}_{\bm{k}'}|^2\} (i\sigma^y)_{\alpha \beta} (i\sigma^y)^\dagger_{\gamma \delta}$ && $A_{1g}$ \\
  && $+ \frac{1}{4} \sum_{n} \sum_{j = x, y} (\bm{c}^{\Gamma n}_{\bm{k}} i\sigma^y)_{\alpha \beta} \cdot (\bm{c}^{\Gamma n}_{\bm{k}'} i\sigma^y)^\dagger_{\gamma \delta}$ && $\Gamma_g \times T_{1g}$ \\
  \\
  OE ($\Gamma_u^+$) && $- \frac{1}{8} \sum_{n} \{|\bm{d}^{\Gamma n}_{\bm{k}}|^2 + |\bm{d}^{\Gamma n}_{\bm{k}'}|^2\} (i\sigma^y)_{\alpha \beta} (i\sigma^y)^\dagger_{\gamma \delta}$ && $A_{1g}$ \\
  && $- \frac{1}{4} \sum_{n} (\bm{d}^{\Gamma n}_{\bm{k}} \cdot \bm{\sigma} i\sigma^y)_{\alpha \beta} (\bm{d}^{\Gamma n}_{\bm{k}'} \cdot \bm{\sigma} i\sigma^y)^\dagger_{\gamma \delta}$ && $\Gamma_u$ \\
  && $+ \frac{1}{4} \sum_{n} \{\bm{d}^{\Gamma n}_{\bm{k}} \times \bm{\sigma} i\sigma^y\}_{\alpha \beta} \cdot \{\bm{d}^{\Gamma n}_{\bm{k}'} \times \bm{\sigma} i\sigma^y\}^\dagger_{\gamma \delta}$ && $\Gamma_u \times T_{1g}$ \\
  \\
  OM ($\Gamma_u^-$) && $+ \frac{1}{8} \sum_{n} \{(\phi^{\Gamma n}_{\bm{k}})^2 + (\phi^{\Gamma n}_{\bm{k}'})^2\} (i\sigma^y)_{\alpha \beta} (i\sigma^y)^\dagger_{\gamma \delta}$ && $A_{1g}$ \\
  && $+ \frac{1}{4} \sum_{n} (\phi^{\Gamma n}_{\bm{k}} \bm{\sigma} i\sigma^y)_{\alpha \beta} \cdot (\phi^{\Gamma n}_{\bm{k}'} \bm{\sigma} i\sigma^y)^\dagger_{\gamma \delta}$ && $\Gamma_u \times T_{1g}$ \\ \hline\hline
 \end{tabularx}
\end{table*}

Tables~\ref{tab:vertex_D4h}--\ref{tab:vertex_Oh} represent the complete classification of the multipole-fluctuation-mediated interaction vertex in crystalline systems, within the condition $V_{\bm{k} \pm \bm{k}'} \simeq V_0$.
In the tables, direct products of two representations are shown, e.g., $\Gamma_g \times A_{2g}$.
They originate from the pseudospin degree of freedom $\sigma^j$ ($j = x, y, z$) of the degenerate band.
A component $\sigma^z$ is distinct from the other components $\sigma^x$ and $\sigma^y$ in tetragonal and hexagonal systems; the former belongs to the $A_{2g}$ IR, while the latter two to the $E_g$ ($E_{1g}$) IR in the $D_{4h}$ ($D_{6h}$) point group.
In cubic systems, on the other hand, all the $x$, $y$, and $z$ directions are equivalent.
Thus, $\{\sigma^x, \sigma^y, \sigma^z\}$ are bases of the three-dimensional IR, $T_{1g}$.
The explicit forms of the direct products are shown in Appendix~\ref{ap:direct_product}.

In the decomposition of the vertex (Tables~\ref{tab:vertex_D4h}--\ref{tab:vertex_Oh}), a term with the sign $+$ and $-$ indicates a repulsive and attractive interaction, respectively.
Thus, we can speculate what superconducting symmetry is likely.
Within the speculation, the main conclusion is similar to that we obtained in isotropic systems.
First, electric multipole fluctuations mediate attractive pairing interactions of not only the totally symmetric IR ($A_{1g}$), but also the same IR ($\Gamma$) as the multipole.
Therefore, OE multipole fluctuations such as ferroelectric fluctuations may stabilize odd-parity spin-triplet superconductivity, while the EE multipole fluctuations such as quadrupole fluctuations favor spin-singlet superconductivity.
We show some examples in the next subsection and discuss candidate materials in Sec.~\ref{sec:candidates}.
Second, in the vicinity of the magnetic multipole order, all the pairing channels are \textit{apparently} repulsive.
Then, we might speculate that the superconductivity is unstable.
In some cases, however, an effectively attractive interaction in the $A_{1g}$ channel could be realized, from a term
\begin{equation}
 + \frac{|V_0|}{8} \sum_{n} \{|\bm{c}^{\Gamma n}_{\bm{k}}|^2 + |\bm{c}^{\Gamma n}_{\bm{k}'}|^2\} (i\sigma^y)_{\alpha \beta} (i\sigma^y)^\dagger_{\gamma \delta},
\end{equation}
for EM multipoles, and
\begin{equation}
 + \frac{|V_0|}{8} \sum_{n} \{(\phi^{\Gamma n}_{\bm{k}})^2 + (\phi^{\Gamma n}_{\bm{k}'})^2\} (i\sigma^y)_{\alpha \beta} (i\sigma^y)^\dagger_{\gamma \delta},
\end{equation}
for OM multipoles, because of the symmetry lowering due to the CEF effect.
We propose anisotropic $s$-wave superconductivity based on this mechanism in the following subsections using some examples.

\subsection{Nodeless $s$-wave and $p$-wave superconductivity by ferroelectric ($A_{2u}^+$ or $E_u^+$) fluctuations in $D_{4h}$ systems}
\label{sec:crystalline-D4h_A2up_Eup}
In the previous subsection, the complete classification tables of the vertex function under the CEF have been given.
Now we discuss some interesting examples of unconventional pairings using the results.
For simplicity, let us assume an isotropic (spherical) Fermi surface in the following discussions.
In this situation, the CEF effect is imposed only on the interaction vertex introduced in Tables~\ref{tab:vertex_D4h}--\ref{tab:vertex_Oh}.
Then substituting the vertex into a linearized gap equation (Appendix~\ref{ap:linearized_gap_eq}), the equation is analytically solvable for each IR channel, with the help of (vector) spherical harmonics (Appendix~\ref{ap:vector_spherical_harmonics}).

In this subsection, let us focus on the OE multipole fluctuation and demonstrate how the result in Eq.~\eqref{eq:interaction_vertex_isotropic_ED} is changed under the $D_{4h}$ CEF effect.
As mentioned in Sec.~\ref{sec:crystalline-multipole}, the OE dipoles $\{x, y, z\}$ in isotropic systems split into $\{z\}$ in the $A_{2u}^+$ and $\{x, y\}$ in the $E_u^+$ IR.
According to Table~\ref{tab:multipole_basis_D4h}(OE), the corresponding bases in momentum space are $\{\hat{k}^x \hat{\bm{y}} - \hat{k}^y \hat{\bm{x}}\}$ for $A_{2u}^+$ and $\{\hat{k}^y \hat{\bm{z}} \pm \hat{k}^z \hat{\bm{y}}, \hat{k}^x \hat{\bm{z}} \pm \hat{k}^z \hat{\bm{x}}\}$ for $E_u^+$.

Here we discuss superconductivity accompanied by the electric dipole (ferroelectric) fluctuation.
For the $A_{2u}^+$ fluctuation ($\bm{d}^{A_{2u}^+}_{\bm{k}} = \hat{k}^x \hat{\bm{y}} - \hat{k}^y \hat{\bm{x}}$), the vertex function in Table~\ref{tab:vertex_D4h}(OE) is given by
\begin{widetext}
 \begin{align}
  V_{\alpha \beta \gamma \delta}(\bm{k}, \bm{k}') &= - \frac{|V_0|}{8} \{(\hat{k}^x)^2 + (\hat{k}^y)^2 + (\hat{k}'^x)^2 + (\hat{k}'^y)^2\} (i\sigma^y)_{\alpha \beta} (i\sigma^y)^\dagger_{\gamma \delta} & A_{1g} \notag \\
  & \quad - \frac{|V_0|}{4} \{(\hat{k}^y \sigma^x - \hat{k}^x \sigma^y) i\sigma^y\}_{\alpha \beta} \{(\hat{k}'^y \sigma^x - \hat{k}'^x \sigma^y) i\sigma^y\}^\dagger_{\gamma \delta} & A_{2u} \notag \\
  & \quad + \frac{|V_0|}{4} \{(\hat{k}^x \sigma^x + \hat{k}^y \sigma^y) i\sigma^y\}_{\alpha \beta} \{(\hat{k}'^x \sigma^x + \hat{k}'^y \sigma^y) i\sigma^y\}^\dagger_{\gamma \delta} & A_{1u} \notag \\
  & \quad + \frac{|V_0|}{4} \sum_{j = x, y} \{\hat{k}^j \sigma^z i\sigma^y\}_{\alpha \beta} \{\hat{k}'^j \sigma^z i\sigma^y\}^\dagger_{\gamma \delta} & E_u.
 \end{align}
\end{widetext}
Therefore attractive interaction appears in the spin-triplet $A_{2u}$ channel as well as the $A_{1g}$ channel, which is similar to the result in the isotropic case.

\begin{figure}[tbp]
 \centering
 \includegraphics[height=4cm]{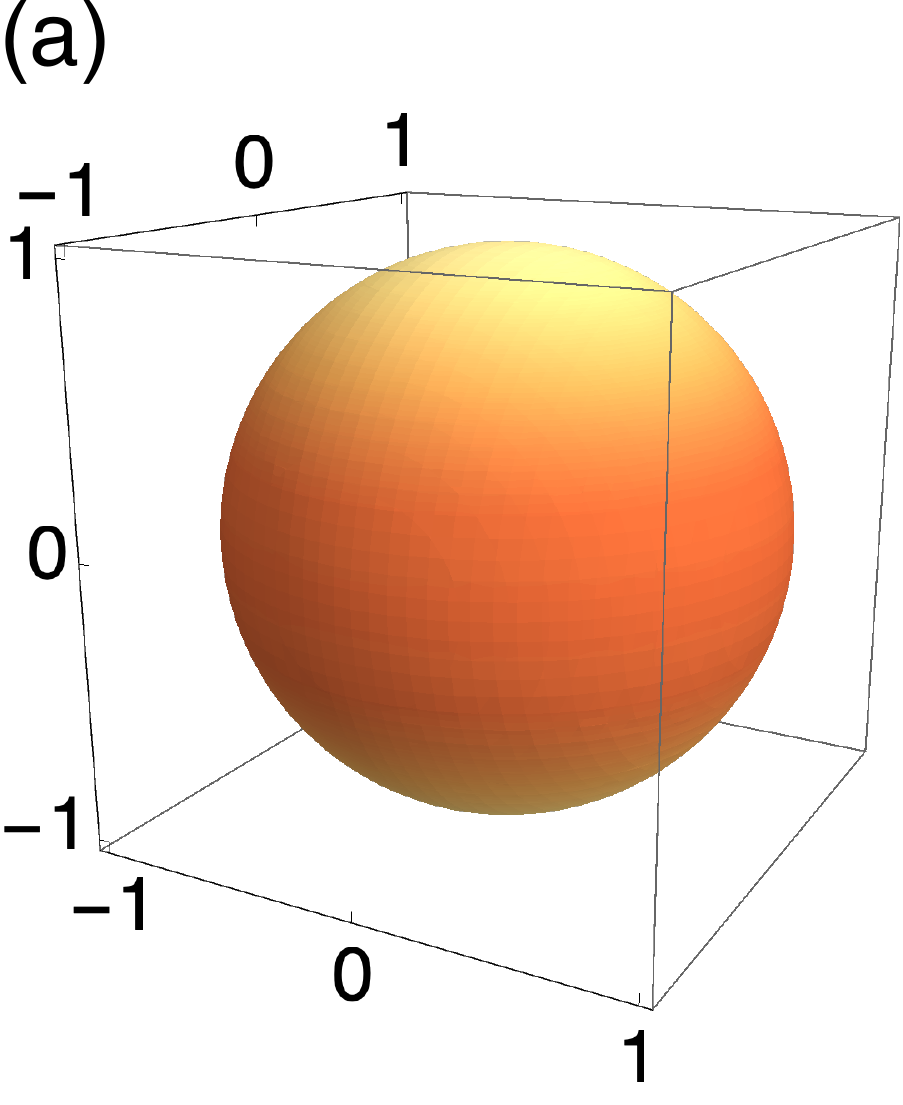}
 \hspace{5mm}
 \includegraphics[height=4cm]{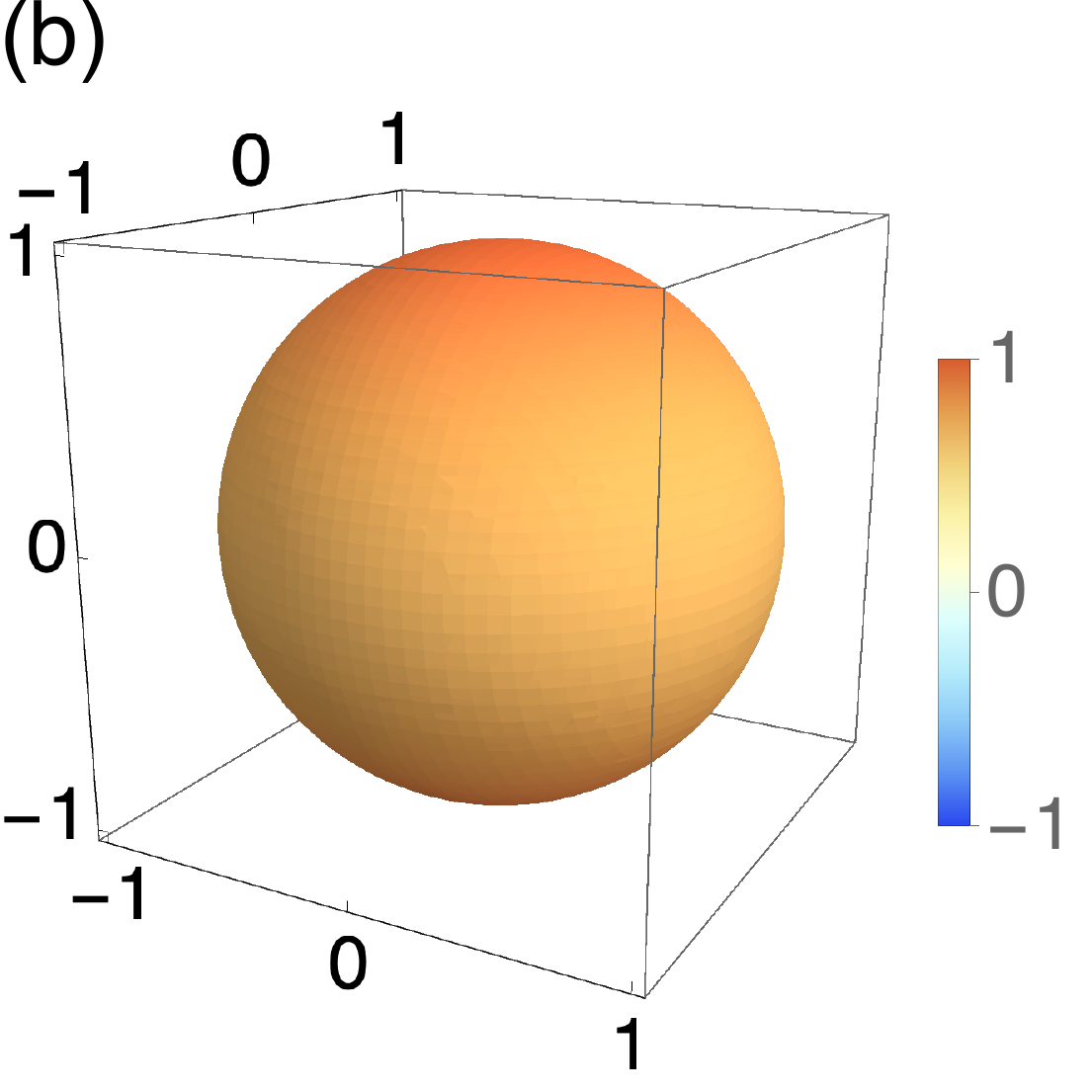}
 \caption{Color map of the nodeless extended $s$-wave order parameter on the spherical Fermi surface induced by the (a) $A_{2u}^+$ and (b) $E_u^+$ multipole fluctuations of $D_{4h}$.}
 \label{fig:extended_s-wave_D4h_A2up_Eup}
\end{figure}

Solving the linearized gap equation (Appendix~\ref{ap:linearized_gap_eq}), we calculate the superconducting transition temperature $T_{\text{c}}$ and the corresponding order parameter.
First, considering the $A_{1g}$ channel, the vertex function is rewritten in spherical harmonics,
\begin{widetext}
 \begin{equation}
  V^{A_{1g}}_{\alpha \beta \gamma \delta}(\bm{k}, \bm{k}') = - 4\pi |V_0| \left[ \frac{1}{3} Y_{0 0}(\hat{\bm{k}}) Y_{0 0}^*(\hat{\bm{k}}') - \frac{1}{6\sqrt{5}} \{Y_{0 0}(\hat{\bm{k}}) Y_{2 0}^*(\hat{\bm{k}}') + Y_{2 0}(\hat{\bm{k}}) Y_{0 0}^*(\hat{\bm{k}}')\} \right] \cdot \frac{1}{2} (i\sigma^y)_{\alpha \beta} (i\sigma^y)^\dagger_{\gamma \delta},
  \label{eq:interaction_vertex_D4h_A2up_A1g}
 \end{equation}
\end{widetext}
where the second term is a nonseparable cross term.
This term appears because of the representation merging under the CEF.
Note that the vertex functions are inevitably separable in the isotropic systems [see Eq.~\eqref{eq:interaction_vertex_isotropic_ED} for example].
We substitute Eq.~\eqref{eq:interaction_vertex_D4h_A2up_A1g} into the linearized gap equation~\eqref{eq:linearized_gap_eq}, and integrate with respect to the solid angle $\Omega_{\hat{\bm{k}}}$ after multiplying $Y_{0 0}^*(\hat{\bm{k}})$ [$Y_{2 0}^*(\hat{\bm{k}})$].
Then, the following simultaneous equations are obtained:
\begin{subequations}
 \label{eq:linearized_gap_eq_A2up_A1g}
 \begin{align}
  c_{0 0} &= N(0) |V_0| \left(\frac{c_{0 0}}{3} - \frac{c_{2 0}}{6\sqrt{5}} \right) \int_{0}^{\omega_c} d\xi \, \frac{1}{\xi} \tanh\left(\frac{\xi}{2T_{\text{c}}^{A_{1g}}}\right), \\
  c_{2 0} &= N(0) |V_0| \left(- \frac{c_{0 0}}{6\sqrt{5}} \right) \int_{0}^{\omega_c} d\xi \, \frac{1}{\xi} \tanh\left(\frac{\xi}{2T_{\text{c}}^{A_{1g}}}\right).
 \end{align}
\end{subequations}
Obviously, both $c_{0 0}$ and $c_{2 0}$ need to be nonzero for the equations possessing a nontrivial solution.
Thus, a quadratic equation about $x = c_{2 0} / c_{0 0}$,
\begin{equation}
 \frac{1}{6\sqrt{5}} x^2 - \frac{1}{3} x - \frac{1}{6\sqrt{5}} = 0,
\end{equation}
is derived.
It has two solutions $x = \sqrt{5} \pm \sqrt{6}$, one of which is negative and satisfies Eq.~\eqref{eq:linearized_gap_eq_A2up_A1g}: $0 > c_{2 0} / c_{0 0} = \sqrt{5} - \sqrt{6}$.
Finally the solution of Eq.~\eqref{eq:linearized_gap_eq_A2up_A1g} gives the following order parameter:
\begin{align}
 \Delta^{A_{1g}}(\bm{k}) &\sim [(\sqrt{30} - 3) \{(\hat{k}^x)^2 + (\hat{k}^y)^2\} \notag \\
 & \qquad + (12 - 2\sqrt{30}) (\hat{k}^z)^2] i\sigma^y,
\end{align}
with a critical temperature
\begin{equation}
 T_{\text{c}}^{A_{1g}} = 1.14 \omega_c \exp\left( - \frac{6\sqrt{30} - 30}{N(0) |V_0|} \right).
 \label{eq:Tc_D4h_A2up_A1g}
\end{equation}
Although the order parameter $\Delta^{A_{1g}}(\bm{k})$ seems to have a complicated form reflecting anisotropy due to the CEF, it indeed represents a slightly distorted $s$-wave gap function as illustrated in Fig.~\ref{fig:extended_s-wave_D4h_A2up_Eup}(a).
Thus, we obtained a solution for the nodeless $s$-wave superconductivity.

For the spin-triplet $A_{2u}$ channel, on the other hand, the vertex function is rewritten in vector spherical harmonics,
\begin{align}
 & V^{A_{2u}}_{\alpha \beta \gamma \delta}(\bm{k}, \bm{k}') \notag \\
 & = - \frac{4\pi}{2} \frac{|V_0|}{3} \{\bm{Y}_{1 0}^{1}(\hat{\bm{k}}) \cdot \bm{\sigma} i\sigma^y\}_{\alpha \beta} \{\bm{Y}_{1 0}^{1}(\hat{\bm{k}}') \cdot \bm{\sigma} i\sigma^y\}^\dagger_{\gamma \delta}.
 \label{eq:interaction_vertex_D4h_A2up_A2u}
\end{align}
Substituting it into the linearized gap equation~\eqref{eq:linearized_gap_eq}, we easily obtain a solution for $p$-wave superconductivity,
\begin{equation}
 \Delta^{A_{2u}}(\bm{k}) \sim \bm{Y}_{1 0}^{1}(\hat{\bm{k}}) \cdot \bm{\sigma} i\sigma^y \sim (\hat{k}^x \sigma^y - \hat{k}^y \sigma^x) i\sigma^y,
\end{equation}
where the $\bm{d}$-vector has the exact same form as the multipole basis $\bm{d}^{A_{2u}^+}_{\bm{k}} = \hat{k}^x \hat{\bm{y}} - \hat{k}^y \hat{\bm{x}}$ in momentum space.
In the $A_{2u}$ superconductivity, point nodes appear on the north and south poles of the Fermi surface.
The existence of the nodal points is ensured by crystal symmetry; a zero-dimensional (0D) $\mathbb{Z}_2$ number defined on the $C_{4v}$-symmetric axis characterizes the nodes~\cite{Sumita2019}.
Furthermore, a two-dimensional (2D) $\mathbb{Z}_2$ topological number~\cite{Sato2009, Sato2010, Fu2010},
\begin{equation}
 \nu_{\text{2D}}^j = \prod_{k^{i \neq j} = 0, \pi; \, k^j = 0} \sgn(\xi_{\bm{k}}) \in \mathbb{Z}_2,
\end{equation}
is well-defined and nontrivial for $j = z$, where $\xi_{\bm{k}}$ is an energy dispersion in the normal state.
Thus, the superconducting state hosts the Majorana mode when we choose an appropriate surface.
The transition temperature $T_{\text{c}}^{A_{2u}}$ of the $p$-wave superconductivity is
\begin{equation}
 T_{\text{c}}^{A_{2u}} = 1.14 \omega_c \exp\left( - \frac{3}{N(0) |V_0|} \right),
 \label{eq:Tc_D4h_A2up_A2u}
\end{equation}
which is slightly lower than the spin-singlet one, $T_{\text{c}}^{A_{1g}}$.

The difference in transition temperatures of $s$-wave and and $p$-wave superconductivity is much smaller than that in the isotropic system~\cite{Kozii2015}.
The logarithm of the ratio of the transition temperatures is
\begin{equation}
 N(0) |V_0| \ln\left(\frac{T_{\text{c}}^{A_{2u}}}{T_{\text{c}}^{A_{1g}}}\right) = (6\sqrt{30} - 30) - 3 = -0.137,
\end{equation}
while that in the isotropic case\footnote{In this case, both transition temperatures are equivalent to those in $O_h$ systems [see $\Gamma = T_{1u}^+$ in Table~\ref{tab:SC_solutions}(c)].} is
\begin{equation}
 N(0) |V_0| \ln\left(\frac{T_{\text{c}}^{\text{$p$-wave}}}{T_{\text{c}}^{\text{$s$-wave}}}\right) = -5.
\end{equation}
Thus, the CEF effect significantly favors $p$-wave superconductivity.
As referred to in Ref.~\cite{Kozii2015}, $T_{\text{c}}$ of the $s$-wave channel is reduced by short-range Coulomb repulsion.
Therefore, the $p$-wave superconductivity may be more stable than the $s$-wave one in nearly ferroelectric \textit{crystalline} systems.
Later we discuss SrTiO$_3$ as a candidate superconductor.

Note that we have adopted the \textit{weak-coupling BCS mean-field theory}.
Therefore, the transition temperatures calculated in Eqs.~\eqref{eq:Tc_D4h_A2up_A1g} and \eqref{eq:Tc_D4h_A2up_A2u} may change when quantum fluctuations of the multipole order parameter are taken into account.
Although it is expected that the symmetry of superconductivity is not altered by the quantum fluctuations in most cases~\cite{Moriya2000, Yanase2003}, it is desirable to refer to higher-order theories when the $s$-wave and $p$-wave states are nearly degenerate.
This is an interesting future issue since we may expect an enhancement of $T_{\text{c}}$ due to the quantum criticality, as Ref.~\cite{Kozii2019} suggested.

\begin{table*}[tbp]
 \centering
 \caption{Solutions of the linearized gap equation~\eqref{eq:linearized_gap_eq} for some multipole-fluctuation-mediated superconductivity in (a) $D_{4h}$, (b) $D_{6h}$, and (c) $O_h$ crystalline systems. When there exist more than one solution for a multipole fluctuation, they are listed in descending order of $T_{\text{c}}$. ``MO'' in (b) and ``EQ'' in (c) represent the momentum-based magnetic octupole [Eq.~\eqref{eq:EM_multipole_D6h_MO}] and electric quadrupole [Eq.~\eqref{eq:EE_multipole_Oh_EQ}], respectively. The seventh column represents topological numbers with one or higher dimensions that have the possibility of being finite (nontrivial) in the superconducting state.}
 \label{tab:SC_solutions}
 \begin{tabularx}{\linewidth}{cXcXcXlXlXcXc} \hline\hline
  $\Gamma$ && $\hat{\bm{k}}$-based multipole && IR && \multicolumn{1}{c}{$T_{\text{c}}$} && \multicolumn{1}{c}{$\Delta(\bm{k})$} && Fig. && Topo. \# \\ \hline
  \multicolumn{13}{c}{(a) Tetragonal ($D_{4h}$)} \\
  $A_{2u}^+$ && $\hat{k}^x \hat{\bm{y}} - \hat{k}^y \hat{\bm{x}}$ && $A_{1g}$ && $1.14 \omega_c \exp\left( - \frac{6\sqrt{30} - 30}{N(0) |V_0|} \right)$ && $[(\sqrt{30} - 3) \{(\hat{k}^x)^2 + (\hat{k}^y)^2\} + (12 - 2\sqrt{30}) (\hat{k}^z)^2] i\sigma^y$ && \ref{fig:extended_s-wave_D4h_A2up_Eup}(a) && N/A \\
  && && $A_{2u}$ && $1.14 \omega_c \exp\left( - \frac{3}{N(0) |V_0|} \right)$ && $(\hat{k}^x \sigma^y - \hat{k}^y \sigma^x) i\sigma^y$ && && $\nu_{\text{2D}}^j$ \\
  $E_u^+$ && $\{\hat{k}^y \hat{\bm{z}} \pm \hat{k}^z \hat{\bm{y}},$ && $A_{1g}$ && $1.14 \omega_c \exp\left( - \frac{6 (\sqrt{105} - 10)}{N(0) |V_0|} \right)$ && $[(12 - \sqrt{105}) \{(\hat{k}^x)^2 + (\hat{k}^y)^2\}$ && \ref{fig:extended_s-wave_D4h_A2up_Eup}(b) && N/A \\
  && $\hat{k}^x \hat{\bm{z}} \pm \hat{k}^z \hat{\bm{x}}\}$ && && && $\qquad\qquad{} + (-18 + 2\sqrt{105}) (\hat{k}^z)^2] i\sigma^y$ && && \\
  && && $E_u$ && $1.14 \omega_c \exp\left( - \frac{3\sqrt{2}}{N(0) |V_0|} \right)$ && $\{(\sqrt{2} - 1) (\hat{k}^x + i\hat{k}^y) \sigma^z \pm \hat{k}^z (\sigma^x + i\sigma^y),$ && && $\nu_{\text{2D}}^j$ or $C_{\text{2D}}$ \\
  && && && && $\,\, (\sqrt{2} - 1) (\hat{k}^x - i\hat{k}^y) \sigma^z \pm \hat{k}^z (\sigma^x - i\sigma^y)\} i\sigma^y$ && && \\
  $A_{2u}^-$ && $\hat{k}^z$ && $A_{1g}$ && $1.14 \omega_c \exp\left( - \frac{9\sqrt{5} + 15}{N(0) |V_0|} \right)$ && $[(\sqrt{5} + 3) \{(\hat{k}^x)^2 + (\hat{k}^y)^2\} - 2 (\sqrt{5} + 1) (\hat{k}^z)^2] i\sigma^y$ && \ref{fig:extended_s-wave_D4h_A2um_Eum}(a) && $w_{\text{1D}}$ \\
  $E_u^-$ && $\{\hat{k}^x, \hat{k}^y\}$ && $A_{1g}$ && $1.14 \omega_c \exp\left( - \frac{6\sqrt{30} + 30}{N(0) |V_0|} \right)$ && $[(\sqrt{30} + 3) \{(\hat{k}^x)^2 + (\hat{k}^y)^2\} - (12 + 2\sqrt{30}) (\hat{k}^z)^2] i\sigma^y$ && \ref{fig:extended_s-wave_D4h_A2um_Eum}(b) && $w_{\text{1D}}$ \\
  \\
  \multicolumn{13}{c}{(b) Hexagonal ($D_{6h}$)} \\
  $E_{1g}^-$ && $\{\hat{\bm{x}}, \hat{\bm{y}}\}$ && \multicolumn{3}{l}{no solution} && && \\
  && MO && $A_{1g}$ && $1.14 \omega_c \exp\left( - \frac{534.451}{N(0) |V_0|} \right)$ && $\{Y_{0 0}(\hat{\bm{k}}) - 0.478411 Y_{2 0}(\hat{\bm{k}}) - 38.1751 Y_{4 0}(\hat{\bm{k}})\} i\sigma^y$ && \ref{fig:extended_s-wave_D6h_E1gm_Oh_Egp}(a) && $w_{\text{1D}}$ \\
  \\
  \multicolumn{13}{c}{(c) Cubic ($O_h$)} \\
  $E_g^+$ && EQ && $E_g$ && $1.14 \omega_c \exp\left( - \frac{10}{N(0) |V_0|} \right)$ && $\{2 (\hat{k}^z)^2 - (\hat{k}^x)^2 - (\hat{k}^y)^2,$ && && $C_{\text{2D}}$ \\
  && && && && $\,\, \sqrt{3} ((\hat{k}^x)^2 - (\hat{k}^y)^2)\} i\sigma^y$ && && \\
  && && $A_{1g}$ && $1.14 \omega_c \exp\left( - \frac{13.8864}{N(0) |V_0|} \right)$ && $[Y_{0 0}(\hat{\bm{k}}) + 0.694319 Y_{4 0}(\hat{\bm{k}})$ && \ref{fig:extended_s-wave_D6h_E1gm_Oh_Egp}(b) && N/A \\
  && && && && $\qquad\qquad{} + 0.414935 \{Y_{4 4}(\hat{\bm{k}}) + Y_{4 -4}(\hat{\bm{k}})\}] i\sigma^y$ && && \\
  $T_{1u}^+$ && $\{(\hat{\bm{k}} \times \bm{\sigma})^n\}$ && $A_{1g}$ && $1.14 \omega_c \exp\left( - \frac{1}{N(0) |V_0|} \right)$ && $i\sigma^y$ && && N/A \\
  && && $T_{1u}$ && $1.14 \omega_c \exp\left( - \frac{6}{N(0) |V_0|} \right)$ && $\{(\hat{\bm{k}} \times \bm{\sigma})^n i\sigma^y\}$ && && $\nu_{\text{2D}}^j$ or $C_{\text{2D}}$ \\ \hline\hline
 \end{tabularx}
\end{table*}

For the $E_u^+$ fluctuations ($\{\bm{d}^{E_u^+ n}_{\bm{k}}\} = \{\hat{k}^y \hat{\bm{z}} \pm \hat{k}^z \hat{\bm{y}}, \hat{k}^x \hat{\bm{z}} \pm \hat{k}^z \hat{\bm{x}}\}$), we can derive possible order parameters and the corresponding transition temperature in a similar way to the above calculation.
The obtained results are shown in Table~\ref{tab:SC_solutions}(a) and Fig.~\ref{fig:extended_s-wave_D4h_A2up_Eup}(b).
The nodeless extended $s$-wave ($A_{1g}$) and the $p$-wave ($E_u$) superconductivity are the most and the second most stable state, respectively.
In the $E_u$ superconductivity, one of the 2D $\mathbb{Z}_2$ numbers $\nu_{\text{2D}}^j$ ($j = x, y, z$) is defined when the superconducting order parameter preserves TRS.
Such order parameter falls into the $B_{2u}$ or $B_{3u}$ IR of the subgroup $D_{2h}$, which has point nodes on the $k_y$ or $k_x$ axis, respectively.
Thus the index $\nu_{\text{2D}}^y$ or $\nu_{\text{2D}}^x$ is correspondingly well-defined and nontrivial.
When TRS is spontaneously broken due to the 2D $E_u$ superconducting order, on the other hand, a 2D Chern number
\begin{equation}
 C_{\text{2D}} = \frac{1}{2 \pi} \oint_S d\bm{S} \cdot \bm{F}(\bm{k}) \in \mathbb{Z},
\end{equation}
can be defined on a closed surface $S$.
Here the Berry flux is defined by using the wave functions whose energy eigenvalue is negative:
\begin{equation}
 F^i(\bm{k}) = - i \varepsilon^{ijk} \sum_{E_n(\bm{k}) < 0} \partial_{k^j} \braket{u_n(\bm{k}) | \partial_{k^k} u_n(\bm{k})}.
\end{equation}
When the $E_u$ superconductivity has a point-nodal gap structure, the 2D index $C_{\text{2D}}$ is nontrivial for $S$ surrounding the node; namely, the point node is a Weyl node.

In contrast to the case of the $A_{2u}^+$ multipole, the functional form of the $p$-wave order parameter induced by the $E_u^+$ multipole fluctuations is deformed from the original multipole bases $\{\bm{d}^{E_u^+ n}_{\bm{k}}\}$ [see Table~\ref{tab:SC_solutions}(a) for an explicit form].
That is because the second attractive term and the third repulsive term in Table~\ref{tab:vertex_D4h}(OE),
\begin{align}
 & - \frac{1}{4} \sum_{n} (\bm{d}^{E_u^+ n}_{\bm{k}} \cdot \bm{\sigma} i\sigma^y)_{\alpha \beta} (\bm{d}^{E_u^+ n}_{\bm{k}'} \cdot \bm{\sigma} i\sigma^y)^\dagger_{\gamma \delta} \notag \\
 & + \frac{1}{4} \sum_{n} \{(\bm{d}^{E_u^+ n}_{\bm{k}} \times \bm{\sigma})^z i\sigma^y\}_{\alpha \beta} \{(\bm{d}^{E_u^+ n}_{\bm{k}'} \times \bm{\sigma})^z i\sigma^y\}^\dagger_{\gamma \delta},
\end{align}
are mixed with each other due to $E_u \times A_{2g} = E_u$.

\subsection{Nodal $s$-wave superconductivity by magnetic toroidal dipole ($A_{2u}^-$ or $E_u^-$) fluctuations in $D_{4h}$ systems}
\label{sec:crystalline-D4h_A2um_Eum}
Next, we discuss fluctuations of OM toroidal dipole moments in $D_{4h}$ superconductors: $y \hat{\bm{x}} - x \hat{\bm{y}}$ for the $A_{2u}^-$ and $\{y \hat{\bm{z}} - z \hat{\bm{y}}, z \hat{\bm{x}} - x \hat{\bm{z}}\}$ for the $E_u^-$ IR.
According to Table~\ref{tab:multipole_basis_D4h}(OM), the corresponding bases in momentum space are $\phi^{A_{2u}^-}_{\bm{k}} = \hat{k}^z$ and $\{\phi^{E_u^- n}_{\bm{k}}\} = \{\hat{k}^x, \hat{k}^y\}$, respectively.

\begin{figure}[tbp]
 \centering
 \includegraphics[height=4cm]{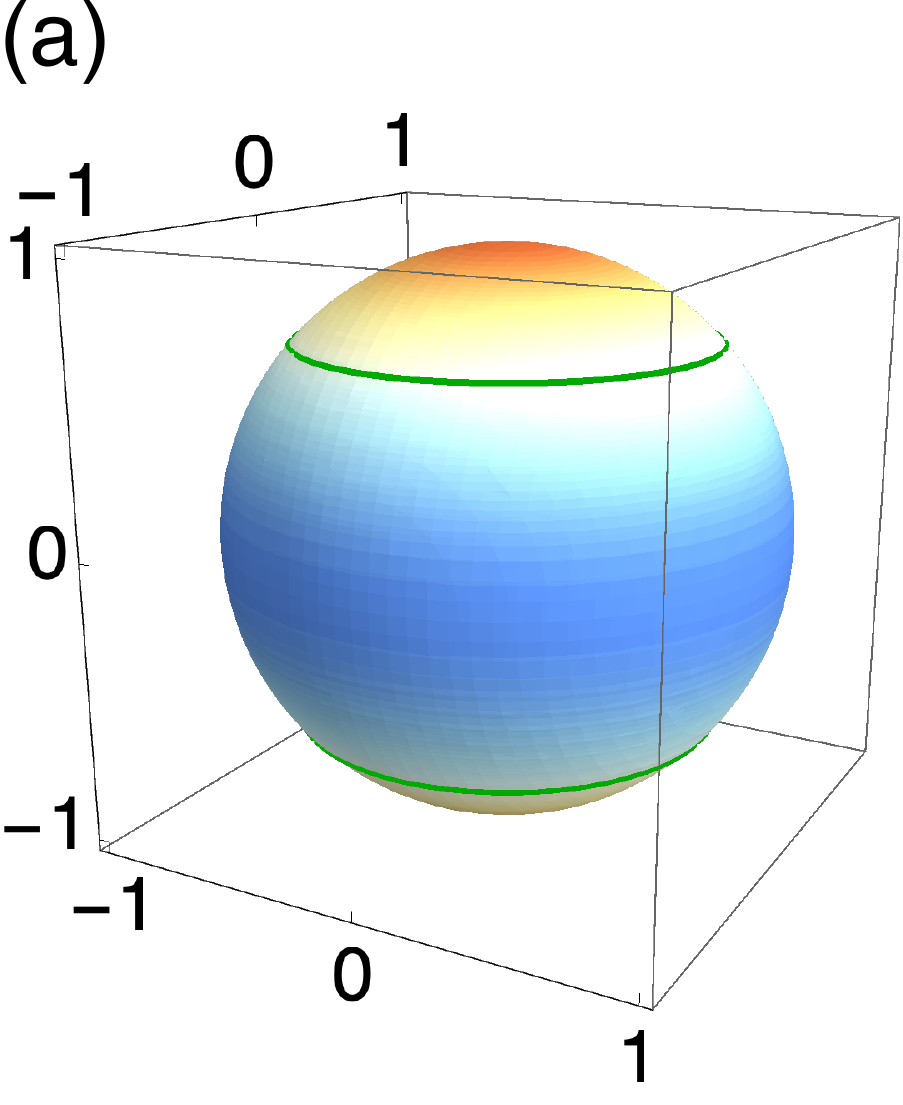}
 \hspace{5mm}
 \includegraphics[height=4cm]{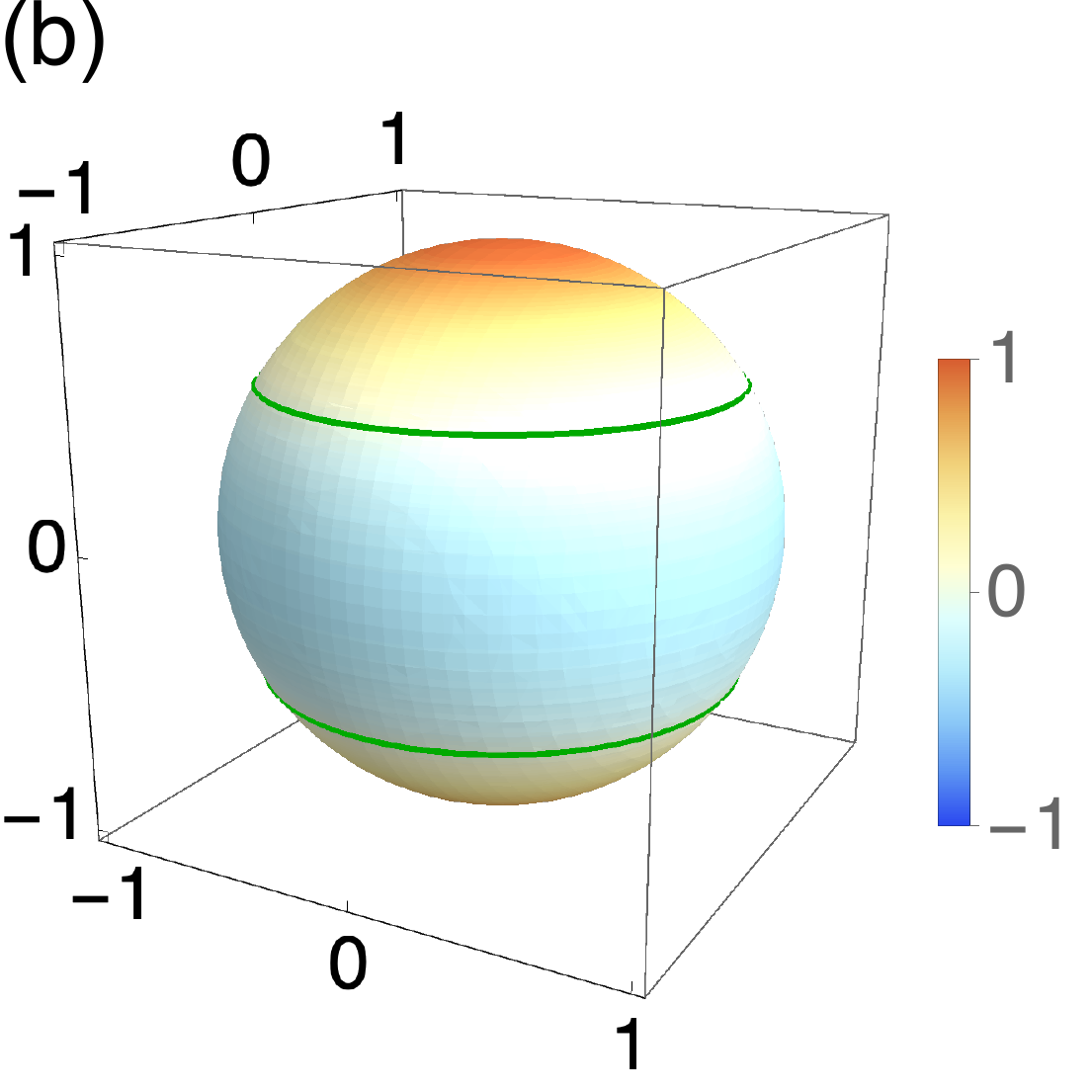}
 \caption{Color map of the nodal extended $s$-wave order parameter on the spherical Fermi surface induced by the (a) $A_{2u}^-$ and (b) $E_u^-$ multipole fluctuations of $D_{4h}$. The green lines represent nodal rings.}
 \label{fig:extended_s-wave_D4h_A2um_Eum}
\end{figure}

Considering the OM fluctuations, the linearized gap equation~\eqref{eq:linearized_gap_eq} has only one solution with a \textit{nodal} extended $s$-wave order parameter, for both $A_{2u}^-$ and $E_u^-$ multipoles [see Table~\ref{tab:SC_solutions}(a) and Fig.~\ref{fig:extended_s-wave_D4h_A2um_Eum}].
Recall that such OM multipoles mediate no attractive pairing in isotropic systems [Eq.~\eqref{eq:interaction_vertex_isotropic_MTD}].
Therefore, the CEF effect plays an essential role to stabilize superconductivity in OM-multipole-fluctuating systems, due to its degeneracy splitting.
An intuitive understanding of the result is shown below.

For concreteness, we consider the $A_{2u}^-$ fluctuation with $\phi^{A_{2u}^-}_{\bm{k}} = \hat{k}^z$; the similar discussion holds for the $E_u^-$ fluctuation.
In this case, the vertex function in Table~\ref{tab:vertex_D4h}(OM) is given by
\begin{align}
 & V_{\alpha \beta \gamma \delta}(\bm{k}, \bm{k}') \notag \\
 & = + \frac{|V_0|}{8} \{(\hat{k}^z)^2 + (\hat{k}'^z)^2\} (i\sigma^y)_{\alpha \beta} (i\sigma^y)^\dagger_{\gamma \delta} & A_{1g} \notag \\
 & \quad + \frac{|V_0|}{4} (\hat{k}^z \sigma^z i\sigma^y)_{\alpha \beta} (\hat{k}'^z \sigma^z i\sigma^y)^\dagger_{\gamma \delta} & A_{1u} \notag \\
 & \quad + \frac{|V_0|}{4} \sum_{j = x, y} (\hat{k}^z \sigma^j i\sigma^y)_{\alpha \beta} (\hat{k}'^z \sigma^j i\sigma^y)^\dagger_{\gamma \delta} & E_u,
\end{align}
where the first $A_{1g}$ term is composed of the summation $(\hat{k}^z)^2 + (\hat{k}'^z)^2$, while the others have separable form between $\hat{\bm{k}}$ and $\hat{\bm{k}}'$.
Then, we rewrite the first term as
\begin{align}
 & V^{A_{1g}}_{\alpha \beta \gamma \delta}(\bm{k}, \bm{k}') \notag \\
 & = + \frac{|V_0|}{8} (i\sigma^y)_{\alpha \beta} (i\sigma^y)^\dagger_{\gamma \delta} \notag \\
 & \quad \times \Bigl[ 1 + (\hat{k}^z)^2 (\hat{k}'^z)^2 \underline{- \{(\hat{k}^x)^2 + (\hat{k}^y)^2\} \{(\hat{k}'^x)^2 + (\hat{k}'^y)^2\}} \Bigr].
\end{align}
The underlined term has a negative sign; namely, an \textit{effectively attractive interaction} is induced in the $A_{1g}$ channel, which results in the extended $s$-wave superconductivity illustrated in Fig.~\ref{fig:extended_s-wave_D4h_A2um_Eum}(a).
Later we discuss hypothetical superconductivity in OM multipole materials.

It is noteworthy that the line nodes illustrated in Figs.~\ref{fig:extended_s-wave_D4h_A2um_Eum}(a) and \ref{fig:extended_s-wave_D4h_A2um_Eum}(b) are topologically protected.
Let $l$ be a closed path encircling the nodal ring.
Then, a one-dimensional (1D) winding number defined by
\begin{equation}
 w_{\text{1D}} = \frac{i}{4\pi} \oint_l d\bm{k} \cdot \Tr[\mathcal{S} \hat{H}_{\text{BdG}}(\bm{k})^{- 1} \nabla_{\bm{k}} \hat{H}_{\text{BdG}}(\bm{k})] \in \mathbb{Z},
 \label{ch3_eq:1D_winding}
\end{equation}
where $\hat{H}_{\text{BdG}}(\bm{k})$ is a Bogoliubov-de Gennes (BdG) Hamiltonian and $\mathcal{S}$ is a chiral operator, has a finite (nontrivial) value $\pm 2$~\cite{Kobayashi2014, Kobayashi2018}.
Therefore, the nodal rings are topologically stable.

\subsection{$s + g$-wave superconductivity by magnetic octupole ($E_{1g}^-$) fluctuations in $D_{6h}$ systems}
\label{sec:crystalline-D6h_E1gm}
Now we focus on a fluctuation of EM multipoles with $E_{1g}^-$ symmetry in hexagonal ($D_{6h}$) superconductors.
The lowest-order bases of the IR are magnetic dipoles $\{\bm{c}^{E_{1g}^- n}_{\bm{k}}\} = \{\hat{\bm{x}}, \hat{\bm{y}}\}$ [Table~\ref{tab:multipole_basis_D4h}(EM)].
For the bases, the interaction vertex function in Table~\ref{tab:vertex_D6h}(EM) is
\begin{equation}
 V_{\alpha \beta \gamma \delta}(\bm{k}, \bm{k}') = + |V_0| (i\sigma^y)_{\alpha \beta} (i\sigma^y)^\dagger_{\gamma \delta} \qquad A_{1g},
 \label{eq:interaction_vertex_D6h_MD}
\end{equation}
which is similar to the isotropic case in Eq.~\eqref{eq:interaction_vertex_isotropic_MD},\footnote{The only difference is the coefficients $3|V_0| / 2$ for isotropic systems and $|V_0|$ for hexagonal systems. It is attributed to the number of components in the magnetic dipole bases taken into account: $\{\hat{\bm{x}}, \hat{\bm{y}}, \hat{\bm{z}}\}$ for the former and $\{\hat{\bm{x}}, \hat{\bm{y}}\}$ for the latter.} and does not result in superconductivity.

\begin{figure}[tbp]
 \centering
 \includegraphics[height=4cm]{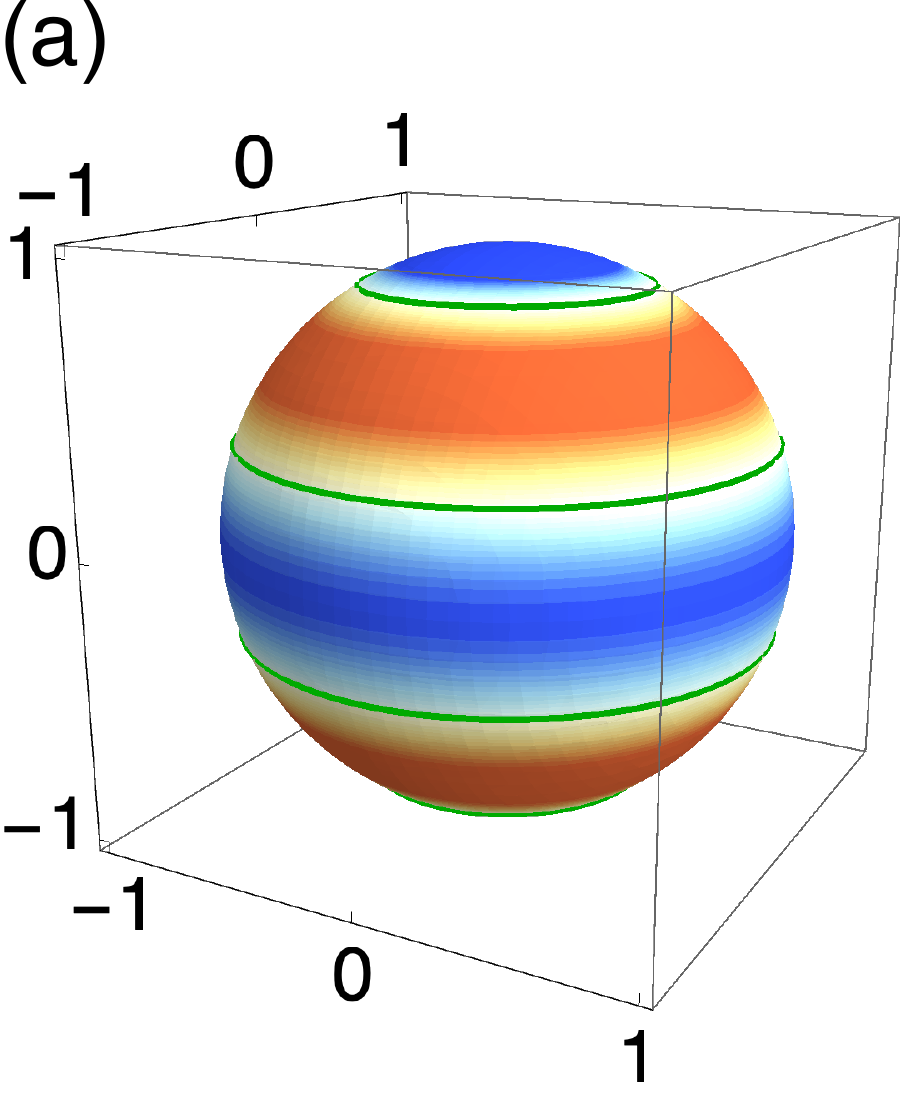}
 \hspace{5mm}
 \includegraphics[height=4cm]{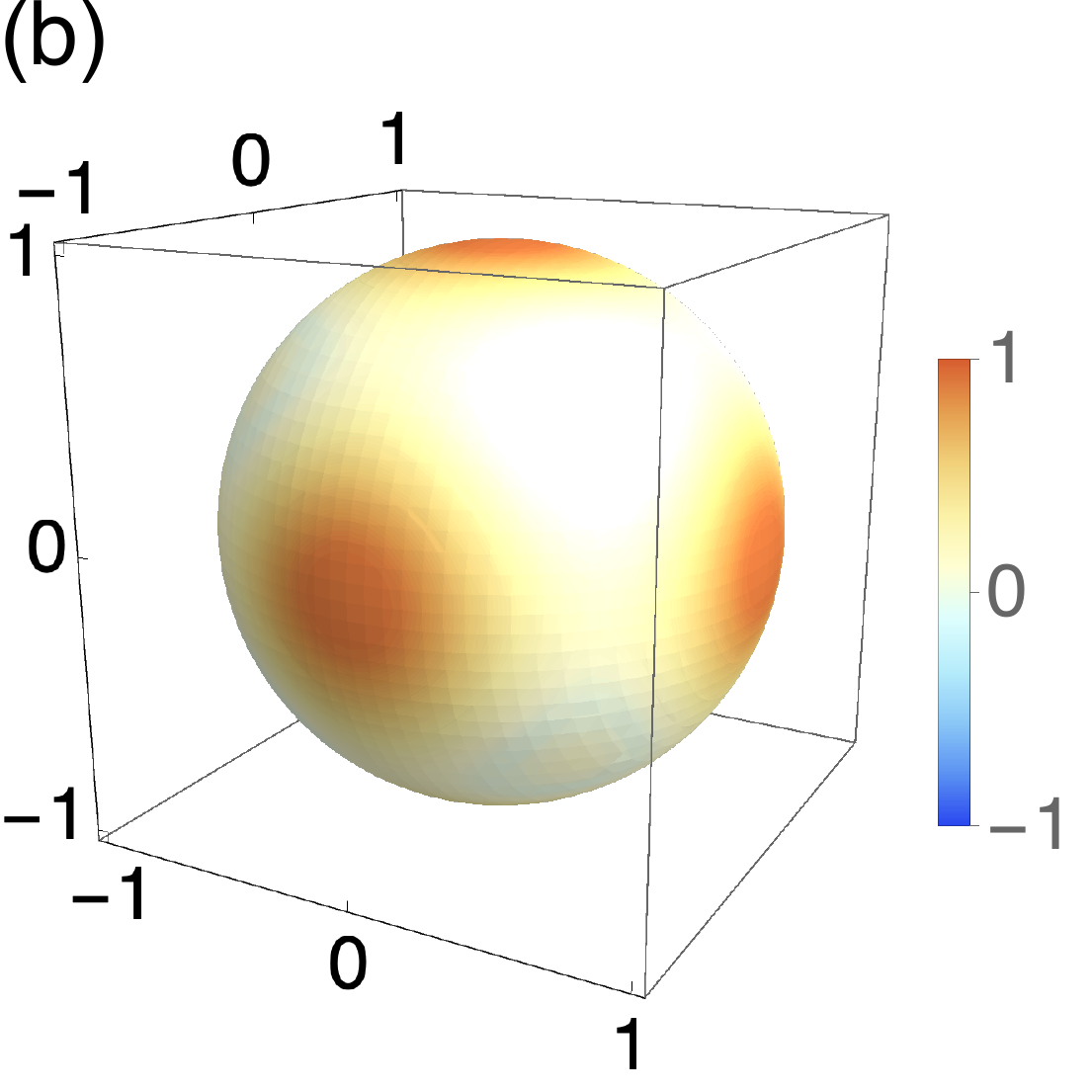}
 \caption{Color map of the $s + g$-wave order parameter on the spherical Fermi surface induced by the (a) $E_{1g}^-$ multipole fluctuations of $D_{6h}$ and (b) $E_g^+$ multipole fluctuations of $O_h$. The green lines represent nodal rings. In (a) $g$-wave Cooper pairing is dominant, while $s$-wave and $g$-wave Cooper pairings are comparable in (b).}
 \label{fig:extended_s-wave_D6h_E1gm_Oh_Egp}
\end{figure}

On the other hand, the $E_{1g}^-$ IR also contains higher-order magnetic octupoles~\cite{Watanabe2018, Hayami2018},
\begin{align}
 & \{\bm{c}^{E_{1g}^- n}_{\bm{k}}\} \notag \\
 & = - \frac{\sqrt{6}}{4} \Bigl\{ \left(3 (\hat{k}^x)^2 + (\hat{k}^y)^2 - 4 (\hat{k}^z)^2\right) \hat{\bm{x}} + 2 \hat{k}^x \hat{k}^y \hat{\bm{y}} - 8 \hat{k}^x \hat{k}^z \hat{\bm{z}}, \notag \\
 & \qquad\qquad 2 \hat{k}^x \hat{k}^y \hat{\bm{x}} + \left((\hat{k}^x)^2 + 3 (\hat{k}^y)^2 - 4 (\hat{k}^z)^2\right) \hat{\bm{y}} - 8 \hat{k}^y \hat{k}^z \hat{\bm{z}} \Bigr\}.
 \label{eq:EM_multipole_D6h_MO}
\end{align}
Using the bases, we get the vertex with three channels: $A_{1g}$, $E_{1g}$, and $E_{2g}$.
When it is substituted into the linearized gap equation~\eqref{eq:linearized_gap_eq}, a single solution with $A_{1g}$ symmetry, which is shown in Table~\ref{tab:SC_solutions}(b), is obtained.
The solution is, as a result of the CEF effect, a highly anisotropic $s + g$-wave order parameter with dominant $g$-wave pairing and four horizontal line nodes [Fig.~\ref{fig:extended_s-wave_D6h_E1gm_Oh_Egp}(a)], which are topologically stable nodes characterized by the 1D winding number $w_{\text{1D}}$.
However, the pairing may be suppressed by the competition with the magnetic dipole fluctuation [Eq.~\eqref{eq:interaction_vertex_D6h_MD}], since $T_{\text{c}}$ is considerably small [Table~\ref{tab:SC_solutions}(b)].
Later we discuss Mn$_3$\textit{Z} (\textit{Z}~$=$~Sn, Ge) as a candidate material.

\subsection{$d$-wave and $s + g$-wave superconductivity by electric quadrupole ($E_g^+$) fluctuations in $O_h$ systems}
\label{sec:crystalline-Oh_Egp}
Let us move on to the cubic ($O_h$) superconductors.
We firstly consider electric quadrupole moments with $E_g^+$ symmetry [Table~\ref{tab:multipole_basis_Oh}(EE)],
\begin{align}
 \{\psi^{E_g^+ n}_{\bm{k}}\} = \biggl\{ & \frac{1}{2} \left(2 (\hat{k}^z)^2 - (\hat{k}^x)^2 - (\hat{k}^y)^2\right), \notag \\
 & \frac{\sqrt{3}}{2} \left((\hat{k}^x)^2 - (\hat{k}^y)^2\right) \biggr\}.
 \label{eq:EE_multipole_Oh_EQ}
\end{align}
For a fluctuation of the quadrupoles, the vertex function in Table~\ref{tab:vertex_Oh}(EE) is given by
\begin{align}
 & V_{\alpha \beta \gamma \delta}(\bm{k}, \bm{k}') \notag \\
 & = - \frac{|V_0|}{8} \Bigl[ \{ (\hat{k}^x)^4 + (\hat{k}^y)^4 + (\hat{k}^z)^4 \notag \\
 & \qquad - (\hat{k}^x)^2 (\hat{k}^y)^2 - (\hat{k}^y)^2 (\hat{k}^z)^2 - (\hat{k}^z)^2 (\hat{k}^x)^2 \} \notag \\
 & \qquad + \{\hat{\bm{k}} \to \hat{\bm{k}}'\} \Bigr] (i\sigma^y)_{\alpha \beta} (i\sigma^y)^\dagger_{\gamma \delta} & A_{1g} \notag \\
 & \quad - \frac{|V_0|}{4} \sum_{n = 1, 2} (\psi^{E_g^+ n}_{\bm{k}} i\sigma^y)_{\alpha \beta} (\psi^{E_g^+ n}_{\bm{k}'} i\sigma^y)^\dagger_{\gamma \delta} & E_g,
\end{align}
which consists of two attractive pairing channels with $A_{1g}$ and $E_g$ symmetry.
Solutions of the linearized gap equation~\eqref{eq:linearized_gap_eq} for the interaction vertex are given in Table~\ref{tab:SC_solutions}(c).
The $d$-wave pairing with the same symmetry as that of the fluctuating electric quadrupoles holds the highest $T_{\text{c}}$, while the anisotropic $s + g$-wave superconductivity is a subleading order.
The $d$-wave superconductivity with $E_g$ symmetry hosts nodes\footnote{Although there exist only point nodes in our minimal single-band model, the nodal points are inflated to surface nodes (Bogoliubov Fermi surfaces) in real superconductors with non-negligible interband pairings~\cite{Agterberg2017, Brydon2018}.} in the $[111]$ direction, which are characterized by a 0D topological index ($\mathbb{Z}$ or $\mathbb{Z}_2$), and a 2D Chern number $C_{\text{2D}}$ when the superconducting order breaks TRS~\cite{Sumita2018, Sumita2019}.
The $s + g$-wave order parameter has no nodal points as illustrated in Fig.~\ref{fig:extended_s-wave_D6h_E1gm_Oh_Egp}(b), unlike the magnetic-octupole-fluctuation-mediated superconductivity with $D_{6h}$ symmetry [Fig.~\ref{fig:extended_s-wave_D6h_E1gm_Oh_Egp}(a)].
Later we discuss PrTi$_2$Al$_{20}$ as a candidate superconductor.

\subsection{A role of momentum dependence in \texorpdfstring{$V_{\bm{q}}$}{$V_q$}}
\label{sec:crystalline-Vq}
Finally we discuss OE dipoles in the $T_{1u}^+$ IR of $O_h$.
In contrast to Sec.~\ref{sec:crystalline-D4h_A2up_Eup} where ferroelectric order was similarly considered, the degeneracy splitting does not occur.
In momentum space, the multipole order parameter is given by $\{\bm{d}^{T_{1u}^+ n}_{\bm{k}}\} = \{(\hat{\bm{k}} \times \bm{\hat{r}})^n\}$, which is equivalent to the $J = 1$ basis functions in isotropic systems [Eq.~\eqref{eq:OE_multipole_isotropic_J1}].
Thus, the interaction vertex [Table~\ref{tab:vertex_Oh}(OE)] was already calculated in Eq.~\eqref{eq:interaction_vertex_isotropic_ED}, that is,
\begin{align}
 & V_{\alpha \beta \gamma \delta}(\bm{k}, \bm{k}') & \notag \\
 & = - \frac{|V_0|}{2} (i\sigma^y)_{\alpha \beta} (i\sigma^y)^\dagger_{\gamma \delta} & A_{1g} \notag \\
 & \quad + \frac{|V_0|}{3} \{\Lambda_{\text{OE}}^{0 0}(\bm{k}) i\sigma^y\}_{\alpha \beta} \{\Lambda_{\text{OE}}^{0 0}(\bm{k}') i\sigma^y\}^\dagger_{\gamma \delta} & A_{1u} \notag \\
 & \quad - \frac{|V_0|}{8} \sum_{i} \{\Lambda_{\text{OE}}^{1 i}(\bm{k}) i\sigma^y\}_{\alpha \beta} \{\Lambda_{\text{OE}}^{1 i}(\bm{k}') i\sigma^y\}^\dagger_{\gamma \delta} & T_{1u} \notag \\
 & \quad + \frac{|V_0|}{16} \sum_{i, j} \{\Lambda_{\text{OE}}^{2, i j}(\bm{k}) i\sigma^y\}_{\alpha \beta} \{\Lambda_{\text{OE}}^{2, i j}(\bm{k}') i\sigma^y\}^\dagger_{\gamma \delta} & E_u + T_{2u},
\end{align}
where we do not decompose the final term into the two IRs $E_u$ and $T_{2u}$ since both channels are repulsive and have no contribution to superconductivity.
As a calculation result of the linearized gap equation~\eqref{eq:linearized_gap_eq}, the first $A_{1g}$ and the third $T_{1u}$ channels induce the predominant $s$-wave and the subleading $p$-wave superconductivity, respectively [see Table~\ref{tab:SC_solutions}(c)].
In the $T_{1u}$ superconductivity, the $\mathbb{Z}_2$ number $\nu_{\text{2D}}^j$ (the Chern number $C_{\text{2D}}$) may be nontrivial when the TRS is preserved (broken).
The value of the indices depends on the rate of the three components in the order parameter.

Here we comment on a recent theoretical suggestion~\cite{Gastiasoro2020} of $s + g$-wave superconductivity by ferroelectric (electric dipole) fluctuations in cubic systems.
In this study, Gastiasoro \textit{et al.} considered the bosonic propagator describing the ferroelectric fluctuations in the disordered state.
The propagator includes the quadratic term with respect to the momentum representing the cubic anisotropy, which induces an anisotropy in the gap function, namely, an admixture of $s$-wave and $g$-wave pairings.
Although the gap anisotropy seems to be not compatible with our result (the purely $s$-wave pairing), we may reproduce, in the framework of the paper, the $s + g$-wave superconductivity by taking into account higher-order terms in the interaction.
For example, the general form of the interaction vertex for $\{\Lambda^{\Gamma n}(\bm{k})\} = \{\bm{d}^{\Gamma n}_{\bm{k}} \cdot \bm{\sigma}\}$ is
\begin{align}
 & V_{\alpha \beta \gamma \delta}(\bm{k}, \bm{k}') \notag \\
 &= \frac{1}{8} \sum_{n} \Bigl[ - V^{\Gamma n}_{\bm{k} - \bm{k}'} (\bm{d}^{\Gamma n}_{\bm{k}} + \bm{d}^{\Gamma n}_{\bm{k}'}) \cdot \bm{\sigma}_{\alpha \delta} (\bm{d}^{\Gamma n}_{\bm{k}} + \bm{d}^{\Gamma n}_{\bm{k}'}) \cdot \bm{\sigma}_{\beta \gamma} \notag \\
 & \qquad + V^{\Gamma n}_{\bm{k} + \bm{k}'} (\bm{d}^{\Gamma n}_{\bm{k}} - \bm{d}^{\Gamma n}_{\bm{k}'}) \cdot \bm{\sigma}_{\alpha \gamma} (\bm{d}^{\Gamma n}_{\bm{k}} - \bm{d}^{\Gamma n}_{\bm{k}'}) \cdot \bm{\sigma}_{\beta \delta} \Bigr].
\end{align}
Here, we adopt an approximation,
\begin{equation}
 V^{\Gamma n}_{\bm{k} \pm \bm{k}'} \simeq V_0 + V_1 (\hat{k}^n \pm \hat{k}'^n)^2 \quad (n = x, y, z),
\end{equation}
while only the first constant term $V_0$ has been considered in this paper.
For the electric dipoles $\{\bm{d}^{T_{1u}^+ n}_{\bm{k}}\} = \{(\hat{\bm{k}} \times \bm{\hat{r}})^n\}$, fourth-order terms about $\hat{\bm{k}}, \hat{\bm{k}}'$ are admixed in the $A_{1g}$ channel when $V_1 \neq 0$, which may results in an $s + g$-wave pairing.

\section{Candidate materials}
\label{sec:candidates}
In this section, we propose some candidate materials of the multipole-fluctuation-induced unconventional superconductivity.
We discuss a stable superconducting phase, based on the results of previous sections, in doped SrTiO$_3$, PrTi$_2$Al$_{20}$, Li$_2$(Pd,~Pt)$_3$B, and magnetic multipole systems.

\subsection{SrTiO$_3$}
Superconductivity in doped SrTiO$_3$ has received extensive attention for a long time~\cite{Schooley1964, Gastiasoro2020_review}.
One of the reasons is that the superconductivity starts to emerge at an extraordinarily low carrier density on the order of $10^{17}$ cm$^{-3}$~\cite{Lin2013, Lin2014}, where the conventional Migdal-Eliashberg theory is not applicable.
Although various studies have proposed possible origins of the dilute superconductivity~\cite{Takada1980, Ruhman2016, Gorkov2016, Edge2015, Dunnett2018, Wolfle2018, ArceGamboa2018, Kedem2018}, there is no sufficient understanding of the pairing mechanism even now.
Another reason is a quantum paraelectricity in SrTiO$_3$, which prevents a long-range ferroelectric order by its quantum fluctuations~\cite{Muller1979, Rowley2014}.
While pure SrTiO$_3$ is in the vicinity of the ferroelectric critical point, it undergoes the ferroelectric transition under some chemical or physical operations~\cite{Bednorz1984, Itoh1999, Uwe1976, Hemberger1995}.
The two major problems, namely the dilute superconductivity and the ferroelectric quantum criticality, are considered to be related with each other.
Indeed, recent theoretical~\cite{Edge2015} and experimental~\cite{Stucky2016, Rischau2017, Tomioka2019, Herrera2019, Ahadi2019} works have suggested an enhancement of the superconducting transition temperature by the ferroelectric quantum fluctuations.

Stimulated by the above backgrounds, many studies have focused on superconducting properties in the vicinity of the ferroelectric (electric dipole) phase~\cite{Kozii2015, Kozii2019, Gastiasoro2020, Edge2015} or in the coexistent ferroelectric phase~\cite{Kanasugi2018, Kanasugi2019, Russell2019}.
Now we apply our result to SrTiO$_3$.
Since SrTiO$_3$ has the crystal structure of the tetragonal space group $I4/mcm$ ($D_{4h}^{18}$) below 105K due to an antiferrodistortive transition~\cite{Fleury1968, Hayward1999}, there are two ferroelectric modes parallel and perpendicular to the antiferrodistortive rotation axis~\cite{Aschauer2014}, which correspond to the $A_{2u}^+$ and $E_u^+$ multipoles, respectively [see Table~\ref{tab:multipole_basis_D4h}(OE)].
As shown in Sec.~\ref{sec:crystalline-D4h_A2up_Eup} and Table~\ref{tab:SC_solutions}(a), the $A_{1g}$ pairing channel gives the highest $T_{\text{c}}$ for the both multipole fluctuations.
Therefore, the nodeless extended $s$-wave superconductivity illustrated in Fig.~\ref{fig:extended_s-wave_D4h_A2up_Eup} is likely realized in doped SrTiO$_3$ by the ferroelectric fluctuations.
The gap symmetry agrees with recent theoretical reports, which explain the dome of $T_{\text{c}}$ in dilute superconductors such as the LaAlO$_3$/SrTiO$_3$ interface~\cite{Zegrodnik2020_arXiv, Boudjada2020_arXiv}.
When the short-range Coulomb interaction suppresses the $s$-wave superconductivity, the subleading $p$-wave superconductivity can be stabilized.
However, the $p$-wave gap function shown in Table~\ref{tab:SC_solutions}(a) possesses nodal points, and it is incompatible with a recent tunneling experiment~\cite{Swartz2018}.
Note that as mentioned in Sec.~\ref{sec:crystalline-Vq}, a $g$-wave order parameter is admixed through higher-order terms in $V_{\bm{k} \pm \bm{k}'}$, which may give a non-negligible effect on the actual SrTiO$_3$ with anisotropic Fermi surfaces~\cite{Mattheiss1972, vanderMarel2011, Hirayama2012, Khalsa2012, Zhong2013}.

\subsection{PrTi$_2$Al$_{20}$}
PrTi$_2$Al$_{20}$ is a cubic superconductor with the space group $Fd\bar{3}m$ ($O_h^7$)~\cite{Niemann1995}.
Recent experiments on the material have observed ferroic ($\bm{q} = 0$) electric quadrupole ordering at $T_{\text{Q}} = 2.0$K, which is closely related to the nonmagnetic $\Gamma_3$ ($E_g$) doublet ground state~\cite{Onimaru2016_review, Sakai2011, Koseki2011, Ito2011, Sato2012}.
Furthermore, PrTi$_2$Al$_{20}$ shows superconductivity at $T_{\text{c}} = 0.2$K~\cite{Sakai2012}.
The superconducting phase continues to exist against pressure~\cite{Onimaru2016_review, Matsubayashi2012, Matsubayashi2014}; in the high-pressure region, the enhancement of $T_{\text{c}}$, which reaches 1.1K at 8.7GPa, and the gradual drop of $T_{\text{Q}}$ above 6.5GPa have been reported.
The enhancement of $T_{\text{c}}$ coinciding with the suppression of $T_{\text{Q}}$ may indicate that the superconductivity is induced by the electric quadrupole fluctuations, although the coexistence of the two phases was reported for at least $P \lesssim 9.1$GPa~\cite{Matsubayashi2012, Matsubayashi2014}.

Now we discuss the superconducting property mediated by the $E_g^+$ electric quadrupole fluctuations.
According to Sec.~\ref{sec:crystalline-Oh_Egp} and Table~\ref{tab:SC_solutions}(c), the $E_g^+$ fluctuation induces the $d$-wave pairing with the same $E_g$ symmetry, which may compete with the $s + g$-wave pairing [Fig.~\ref{fig:extended_s-wave_D6h_E1gm_Oh_Egp}(b)].
Since the $d$-wave superconductivity has two-component order parameters, two possible ground states are expected.
One is a chiral $d$-wave state with broken TRS, and the other is a nematic state possessing a nodal gap.
On the other hand, the $s + g$-wave state is nodeless as shown in Fig.~\ref{fig:extended_s-wave_D6h_E1gm_Oh_Egp}(b).
Although pairing symmetry of PrTi$_2$Al$_{20}$ has not been determined, these superconducting states can be distinguished by experiments.
Our result paves a way to understand the unconventional superconductivity mediated by the electric quadrupole fluctuations.

Whereas PrTi$_2$Al$_{20}$ is a unique compound exhibiting the ferroquadrupole order in the caged Pr 1-2-20 family, the others with an \textit{antiferro}-quadrupole order (e.g. PrV$_2$Al$_{20}$ and PrIr$_2$Zn$_{20}$) may be dealt with in our framework, by taking into account higher-order terms in $V_{\bm{k} \pm \bm{k}'}$ and/or a nesting of the Fermi surface.
We leave this argument for future works.

\subsection{Li$_2$(Pd,~Pt)$_3$B}
Li$_2$Pd$_3$B and Li$_2$Pt$_3$B are ternary borides with an antiperovskite cubic structure belonging to the \textit{noncentrosymmetric} space group $P4_{3}32$ ($O^6$)~\cite{Eibenstein1997}.
In the whole family of Li$_2$(Pd$_{1-x}$Pt$_x$)$_3$B ($0 \leq x \leq 1$), a bulk superconducting transition with a critical temperature $T_{\text{c}} = 2\text{-}8$K was observed by various experiments~\cite{Nishiyama2005, Nishiyama2007, Harada2010, Harada2012, Takeya2007, Eguchi2013, Yuan2006, Hafliger2009, Togano2004, Shamsuzzaman2010, Peets2011}.
The noncentrosymmetry of Li$_2$(Pd$_{1-x}$Pt$_x$)$_3$B is expected to cause mixing of a spin-singlet state and a spin-triplet state in the superconductivity.
Indeed, NMR~\cite{Nishiyama2005, Nishiyama2007, Harada2010, Harada2012}, specific heat~\cite{Takeya2007, Eguchi2013}, and penetration depth~\cite{Yuan2006} measurements have suggested that a line-nodal state with dominant spin-triplet pairing is realized for $x \gtrsim 0.8$ while spin-singlet-dominant BCS superconductivity occurs for $x \lesssim 0.8$, although a $\mu$SR study~\cite{Hafliger2009} supports $s$-wave superconductivity across the entire doping regime.

One possible origin of the spin-triplet-dominant superconductivity for $x \gtrsim 0.8$ is an abrupt increase of the extent of inversion-symmetry breaking due to a local structural distortion~\cite{Harada2012}.
The abrupt change results in an enhancement of the spin-orbit coupling splitting, which has been confirmed by band calculations~\cite{Harada2012, Lee2005, Shishidou_private}.
Therefore, the origin of the superconductivity in Li$_2$(Pd$_{1-x}$Pt$_x$)$_3$B around $x = 0.8$ can be attributed to an enhancement of a momentum-based OE multipole fluctuation, which is a very subject in this paper.
Since the crystal point group of the compound is $O$, the OE multipole belongs to the $A_{1u}^+$ IR of $O_h$.
According to Table~\ref{tab:multipole_basis_Oh}(OE), a real-space basis function of the $A_{1u}^+$ IR is an electric rank-9 multipole (512-pole) $x y z (x^2 - y^2) (y^2 - z^2) (z^2 - x^2)$, which corresponds to a hedgehog structure $\bm{d}_{\bm{k}}^{A_{1u}^+} = \hat{k}^x \hat{\bm{x}} + \hat{k}^y \hat{\bm{y}} + \hat{k}^z \hat{\bm{z}}$ in momentum space.

Next, we consider a stable superconducting state due to the OE multipole fluctuation.
The momentum basis $\bm{d}_{\bm{k}}^{A_{1u}^+}$ is the same as the $J = 0$ multipole in isotropic systems [Eq.~\eqref{eq:OE_multipole_isotropic_J0}].
Thus, the interaction vertex was already calculated in Ref.~\cite{Kozii2015}, that is,
\begin{align}
 & V_{\alpha \beta \gamma \delta}(\bm{k}, \bm{k}') & \notag \\
 &= - \frac{|V_0|}{4} (i\sigma^y)_{\alpha \beta} (i\sigma^y)^\dagger_{\gamma \delta} & A_{1g} \notag \\
 & \quad - \frac{|V_0|}{4} (\hat{\bm{k}} \cdot \bm{\sigma} i\sigma^y)_{\alpha \beta} (\hat{\bm{k}}' \cdot \bm{\sigma} i\sigma^y)^\dagger_{\gamma \delta} & A_{1u} \notag \\
 & \quad + \frac{|V_0|}{4} (\hat{\bm{k}} \times \bm{\sigma} i\sigma^y)_{\alpha \beta} \cdot (\hat{\bm{k}}' \times \bm{\sigma} i\sigma^y)^\dagger_{\gamma \delta} & T_{1u},
 \label{eq:interaction_vertex_isotropic_ET}
\end{align}
which stabilizes $A_{1g}$ ($s$-wave; the first term) and $A_{1u}$ ($p$-wave; the second term) superconductivity.
In the noncentrosymmetric point group $O$, the $A_{1g}$ and $A_{1u}$ IRs of $O_h$ are merged into the identical IR $A_1$.
Therefore, parity-mixed $s + p$-wave superconductivity may occur in Li$_2$(Pd,~Pt)$_3$B and OE multipole fluctuations actually give rise to attractive interactions in both channels.
Such parity mixing causes the presence of line nodes when the spin-triplet component is larger than the spin-singlet one~\cite{Hayashi2006}.
The result is consistent with the line-nodal spin-triplet-dominant superconductivity for $x \gtrsim 0.8$ observed by many experiments~\cite{Nishiyama2005, Nishiyama2007, Harada2010, Harada2012, Takeya2007, Eguchi2013, Yuan2006}.
A similar enhancement of spin-triplet pairing was recently observed in Cd$_2$Re$_2$O$_7$ in the vicinity of the parity violating structural transition~\cite{Kitagawa2020}.

Now we comment on the $s + p + f$-wave superconductivity of Li$_2$Pt$_3$B proposed in Ref.~\cite{Yuan2006}.
In the framework of this paper, the $f$-wave order parameter stems from a cubic component of $\bm{k}$ in the $\bm{g}$-vector, which represents the antisymmetric spin-orbit coupling and corresponds to the momentum basis of the OE multipole $\bm{d}_{\bm{k}}$ in our formalism.
Thus, the result of Ref.~\cite{Yuan2006} should be reproduced by taking into account the following $\bm{d}_{\bm{k}}$,
\begin{align}
 \bm{d}_{\bm{k}}^{A_{1u}^+} &= a_1 (\hat{k}^x \hat{\bm{x}} + \hat{k}^y \hat{\bm{y}} + \hat{k}^z \hat{\bm{z}}) \notag \\
 & \quad + a_2 [\hat{k}^x \{(\hat{k}^y)^2 + (\hat{k}^z)^2\} \hat{\bm{x}} \notag \\
 & \quad\qquad + \hat{k}^y \{(\hat{k}^z)^2 + (\hat{k}^x)^2\} \hat{\bm{y}} \notag \\
 & \quad\qquad + \hat{k}^z \{(\hat{k}^x)^2 + (\hat{k}^y)^2\} \hat{\bm{z}}].
\end{align}
where both the first linear term and the second cubic term of $\bm{k}$ belong to the same IR $A_{1u}^+$.
The cubic term causes the attractive interaction in the $f$-wave pairing channel.
Since the Fermi surfaces of Li$_2$(Pd,~Pt)$_3$B are highly anisotropic~\cite{Harada2012, Lee2005, Shishidou_private}, the cubic term may not be negligible and results in the $f$-wave pairing.

\subsection{Magnetic multipole systems}
Recent vigorous studies have gotten us to recognize that the multipole expansion of a magnetic structure is a powerful approach in the study of magnets and related phenomena.
A recently developed cluster multipole theory~\cite{Suzuki2017, Suzuki2019}, which systematically characterizes a magnetic structure over atoms as a magnetic multipole, has elucidated that coplanar antiferromagnets Mn$_3$\textit{Z} (\textit{Z}~$=$~Sn, Ge) possess a magnetic octupole order corresponding to Eq.~\eqref{eq:EM_multipole_D6h_MO}.
Although Mn$_3$\textit{Z} is a magnetic metal, we may consider a hypothetical superconducting transition which can be realized by carrier doping or applying pressure, etc.
According to Sec.~\ref{sec:crystalline-D6h_E1gm} and Table~\ref{tab:SC_solutions}(b), we naively expect line-nodal $s + g$-wave superconductivity with dominant $g$-wave pairing [Fig.~\ref{fig:extended_s-wave_D6h_E1gm_Oh_Egp}(a)] due to fluctuations of the magnetic octupole.
A similar anisotropic superconducting state may be realized even in a magnetic octupole ice Ce$_2$Sn$_2$O$_7$ with a pyrochlore (cubic) lattice~\cite{Sibille2020}.

Although the magnetic octupole is an EM multipole, the OM multipole order with broken inversion symmetry and TRS has recently been attracting significant attention.
For example, Fulde-Ferrell-Larkin-Ovchinnikov superconductivity coexisting with the magnetic quadrupole~\cite{Sumita2016, Sumita2017}, magnetoelectric effect~\cite{Spaldin2008, Spaldin2013, Yanase2014, Watanabe2018, Hayami2018, Saito2018} and magnetopiezoelectric effect~\cite{Watanabe2018, Shiomi2019_1, Shiomi2019_2} originating from the odd-parity magnetic multipole have been suggested.
More than 110 OM multipole materials have been identified by a recent group-theoretical study~\cite{Watanabe2018}.
At least more than 40 of them exhibit a metallic or semiconducting property.
Thus, the material list is strongly expected to include OM-multipole-fluctuation-mediated superconductors.
As exemplified in Sec.~\ref{sec:crystalline-D4h_A2um_Eum}, nodal extended $s$-wave superconductivity may be stabilized in such systems with the CEF effect.

\section{Summary and discussion}
\label{sec:summary}
In this paper, we elucidated a pairing interaction between electrons mediated by fluctuations of various ferroic multipole orders, which has been an undiscovered pairing mechanism of unconventional superconductivity.
First, we formulated the interaction vertex for all symmetry classes (EE, EM, OE, and OM) of multipoles [Eq.~\eqref{eq:interaction_vertex_general}].
The formulation is useful for a further investigation of induced superconductivity, as we have actually done in the paper.

Next, considering \textit{isotropic} systems, we showed that electric multipole fluctuations mediate not only an $s$-wave pairing but also an unconventional pairing with the same symmetry as the multipole.
This result is consistent with previous studies~\cite{Kozii2015, Kozii2019}, suggesting odd-parity superconductivity by OE multipole fluctuations.
On the other hand, the vertex arising from magnetic multipole fluctuations reveals repulsive interaction in all channels.
Thus, superconductivity is not stabilized in isotropic systems by the magnetic multipole fluctuations.

Furthermore, we carried out exhaustive calculations of pairing vertex mediated by multipole fluctuations in \textit{crystalline} systems.
Correspondence between symmetries of superconductivity and multipole order was clarified.
The order parameter and transition temperature of superconductivity were calculated for some multipole fluctuations in $D_{4h}$, $D_{6h}$, and $O_h$ point groups.
The results are different from isotropic systems owing to the CEF effects.
For example, nodal extended $s$-wave superconductivity may emerge due to magnetic multipole fluctuations, which is a consequence of the degeneracy splitting under the CEF.
Other interesting results are summarized in Table~\ref{tab:SC_solutions}.

Finally, we proposed doped SrTiO$_3$, PrTi$_2$Al$_{20}$, Li$_2$(Pd,~Pt)$_3$B, and some magnetic multipole systems as candidate materials for the multipole-fluctuation-mediated unconventional superconductivity.
In addition, our result may be consistent with a recent theory~\cite{Yamakawa2017}, which suggested the $T_{\text{c}}$ enhancement for both $d$-wave and extended $s$-wave pairing due to nematic orbital fluctuations in Fe-based superconductors~\cite{Kontani2014, Bohmer2015, Hosoi2016, Massat2016}.
Since the recently confirmed symmetry-based approach~\cite{Watanabe2018, Hayami2018} enables us to easily search for multipole ordering systems, it is strongly expected that many other candidates will be discovered.
Thus, our theory becomes a solid foundation for further investigations of exotic superconductivity in the vicinity of the multipole orders.

\begin{acknowledgments}
 The authors are grateful to M.~Sigrist and S.~Kanasugi for fruitful discussions.
 This work was supported by Grants-in-Aid for Scientific Research on Innovative Areas ``J-Physics'' (No. JP15H05884) and ``Topological Materials Science'' (No. JP16H00991 and No. JP18H04225) from JSPS of Japan, by ``J-Physics: Young Researchers Exchange Program'' (No. JP15K21732), by JSPS KAKENHI Grants No. JP15K05164, No. JP17J09908, No. JP18H05227, and No. JP18H01178, and by JST CREST Grant No. JPMJCR19T2.
\end{acknowledgments}

\appendix
\section{Classification of multipoles under CEF}
\label{ap:classification_multipole}
In Tables~\ref{tab:multipole_basis_D4h}--\ref{tab:multipole_basis_Oh}, we reprint classification of real-space and momentum-space basis functions of multipoles under the CEF effect~\cite{Watanabe2018, Hayami2018}.

\begin{table}[tbp]
 \centering
 \caption{Basis functions in the real and momentum spaces for IRs of the $D_{4h}$ point group~\cite{Watanabe2018}. The bases in momentum space (third column) correspond to $\psi^{\Gamma n}_{\bm{k}}$ (EE), $\bm{c}^{\Gamma n}_{\bm{k}}$ (EM), $\bm{d}^{\Gamma n}_{\bm{k}}$ (OE), and $\phi^{\Gamma n}_{\bm{k}}$ (OM).}
 \label{tab:multipole_basis_D4h}
 \begin{tabularx}{\linewidth}{lXcXc} \hline\hline
  IR ($\Gamma$) && Basis in real space && Basis in momentum space \\ \hline
  \multicolumn{5}{c}{(EE)} \\
  $A_{1g}^+$ && $x^2 + y^2, z^2$ && \\
  $A_{2g}^+$ && $x y (x^2 - y^2)$ && \\
  $B_{1g}^+$ && $x^2 - y^2$ && $(\bm{r} \rightarrow \hat{\bm{k}})$ \\
  $B_{2g}^+$ && $xy$ && \\
  $E_{g}^+$  && $\{yz, zx\}$ && \\
  \\
  \multicolumn{5}{c}{(EM)} \\
  $A_{1g}^-$ && $z (y \hat{\bm{x}} - x \hat{\bm{y}})$ && \\
  $A_{2g}^-$ && $\hat{\bm{z}}$ && \\
  $B_{1g}^-$ && $x y \hat{\bm{z}}, z (y \hat{\bm{x}} + x \hat{\bm{y}})$ && $(\bm{r} \rightarrow \hat{\bm{k}})$ \\
  $B_{2g}^-$ && $(x^2 - y^2) \hat{\bm{z}}, \, z (x \hat{\bm{x}} - y \hat{\bm{y}})$ && \\
  $E_{g}^-$  && $\{\hat{\bm{x}}, \hat{\bm{y}}\}$ && \\
  \\
  \multicolumn{5}{c}{(OE)} \\
  $A_{1u}^+$ && $x y z (x^2 - y^2)$ && $\hat{k}^x \hat{\bm{x}} + \hat{k}^y \hat{\bm{y}} + \hat{k}^z \hat{\bm{z}},$ \\
             && && $2 \hat{k}^z \hat{\bm{z}} - \hat{k}^x \hat{\bm{x}} - \hat{k}^y \hat{\bm{y}}$ \\
  $A_{2u}^+$ && $z, z^3$ && $\hat{k}^x \hat{\bm{y}} - \hat{k}^y \hat{\bm{x}}$ \\
  $B_{1u}^+$ && $x y z$ && $\hat{k}^x \hat{\bm{x}} - \hat{k}^y \hat{\bm{y}}$ \\
  $B_{2u}^+$ && $z (x^2 - y^2)$ && $\hat{k}^x \hat{\bm{y}} + \hat{k}^y \hat{\bm{x}}$ \\
  $E_{u}^+$  && $\{x, y\}$ && $\{\hat{k}^y \hat{\bm{z}} + \hat{k}^z \hat{\bm{y}}, \hat{k}^z \hat{\bm{x}} + \hat{k}^x \hat{\bm{z}}\},$ \\
             && && $\{\hat{k}^y \hat{\bm{z}} - \hat{k}^z \hat{\bm{y}}, \hat{k}^z \hat{\bm{x}} - \hat{k}^x\hat{\bm{z}}\}$ \\
  \\
  \multicolumn{5}{c}{(OM)} \\
  $A_{1u}^-$ && $2z \hat{\bm{z}} - x \hat{\bm{x}} - y \hat{\bm{y}},$ && $\hat{k}^x \hat{k}^y \hat{k}^z ((\hat{k}^x)^2 - (\hat{k}^y)^2)$ \\
             && $x \hat{\bm{x}} + y \hat{\bm{y}} + z \hat{\bm{z}}$ && \\
  $A_{2u}^-$ && $y \hat{\bm{x}} - x \hat{\bm{y}}$ && $\hat{k}^z$ \\
  $B_{1u}^-$ && $x \hat{\bm{x}} - y \hat{\bm{y}}$ && $\hat{k}^x \hat{k}^y \hat{k}^z$ \\
  $B_{2u}^-$ && $y \hat{\bm{x}} + x \hat{\bm{y}}$ && $\hat{k}^z ((\hat{k}^x)^2 - (\hat{k}^y)^2)$ \\
  $E_{u}^-$  && $\{y \hat{\bm{z}} + z \hat{\bm{y}}, z \hat{\bm{x}} + x \hat{\bm{z}}\},$ && \multirow{1}{*}{$\{\hat{k}^x, \hat{k}^y\}$} \\
             && $\{y \hat{\bm{z}} - z \hat{\bm{y}}, z \hat{\bm{x}} - x \hat{\bm{z}}\}$ && \\ \hline\hline
 \end{tabularx}
\end{table}

\begin{table*}[tbp]
 \centering
 \caption{Basis functions in the real and momentum spaces for IRs of the $D_{6h}$ point group~\cite{Watanabe2018}.}
 \label{tab:multipole_basis_D6h}
 \begin{tabularx}{\linewidth}{lXcXc} \hline\hline
  IR ($\Gamma$) && Basis in real space && Basis in momentum space \\ \hline
  \multicolumn{5}{c}{(EE)} \\
  $A_{1g}^+$ && $z^2$ && \\
  $A_{2g}^+$ && $3 x^5 y - 10 x^3 y^3 + 3 x y^5$ && \\
  $B_{1g}^+$ && $(3 x^2 - y^2) y z$ && \\
  $B_{2g}^+$ && $(x^2 - 3 y^2) z x$ && $(\bm{r} \rightarrow \hat{\bm{k}})$ \\
  $E_{1g}^+$ && $\{y z, z x\}$ && \\
  $E_{2g}^+$ && $\{x y, x^2 - y^2\}$ && \\
  \\
  \multicolumn{5}{c}{(EM)} \\
  $A_{1g}^-$ && $z (x \hat{\bm{y}} - y \hat{\bm{x}}) $ && \\
  $A_{2g}^-$ && $\hat{\bm{z}}$ && \\
  $B_{1g}^-$ && $(x^2 - y^2) \hat{\bm{x}} - 2 x y \hat{\bm{y}}$ && \\
  $B_{2g}^-$ && $(x^2 - y^2) \hat{\bm{y}} + 2 x y \hat{\bm{x}}$ && $(\bm{r} \rightarrow \hat{\bm{k}})$ \\
  $E_{1g}^-$ && $\{\hat{\bm{x}}, \hat{\bm{y}}\}$ && \\
  $E_{2g}^-$ && $\{(x^2 - y^2) \hat{\bm{z}}, x y \hat{\bm{z}}\}, \, \{z (x \hat{\bm{x}} - y \hat{\bm{y}}), z (y \hat{\bm{x}} + x \hat{\bm{y}})\}$ && \\
  \\
  \multicolumn{5}{c}{(OE)} \\
  $A_{1u}^+$ && $x y z (3 x^4 - 10 x^2 y^2 + 3 y^4)$ && $\hat{k}^z \hat{\bm{z}}, \hat{k}^x \hat{\bm{x}} + \hat{k}^y \hat{\bm{y}}$ \\
  $A_{2u}^+$ && $z$ && $\hat{k}^x \hat{\bm{y}} - \hat{k}^y \hat{\bm{x}}$ \\
  $B_{1u}^+$ && $x (x^2 - 3 y^2)$ && $\hat{k}^y (3 (\hat{k}^x)^2 - (\hat{k}^y)^2) \hat{\bm{z}}, \, \hat{k}^z ((\hat{k}^x)^2 - (\hat{k}^y)^2) \hat{\bm{y}} + 2 \hat{k}^x \hat{k}^y \hat{k}^z \hat{\bm{x}}$ \\
  $B_{2u}^+$ && $y (3 x^2 - y^2)$ && $\hat{k}^x ((\hat{k}^x)^2 - 3 (\hat{k}^y)^2) \hat{\bm{z}}, \, \hat{k}^z ((\hat{k}^x)^2 - (\hat{k}^y)^2) \hat{\bm{x}} - 2 \hat{k}^x \hat{k}^y \hat{k}^z \hat{\bm{y}}$ \\
  $E_{1u}^+$ && $\{x, y\}$ && $\{\hat{k}^y \hat{\bm{z}} + \hat{k}^z \hat{\bm{y}}, \hat{k}^z \hat{\bm{x}} + \hat{k}^x \hat{\bm{z}}\}, \{\hat{k}^y \hat{\bm{z}} - \hat{k}^z \hat{\bm{y}}, \hat{k}^z \hat{\bm{x}} - \hat{k}^x \hat{\bm{z}}\}$ \\
  $E_{2u}^+$ && $\{(x^2 - y^2) z, x y z\}$ && $\{\hat{k}^x \hat{\bm{y}} + \hat{k}^y \hat{\bm{x}}, \hat{k}^x \hat{\bm{x}} - \hat{k}^y \hat{\bm{y}}\}$ \\
  \\
  \multicolumn{5}{c}{(OM)} \\
  $A_{1u}^-$ && $2 z \hat{\bm{z}} - x \hat{\bm{x}} - y \hat{\bm{y}}, x \hat{\bm{x}} + y \hat{\bm{y}} + z\hat{\bm{z}}$ && \multirow{1}{*}{$\hat{k}^x \hat{k}^y \hat{k}^z (3 (\hat{k}^x)^4 - 10 (\hat{k}^x)^2 (\hat{k}^y)^2 + 3 (\hat{k}^y)^4)$} \\
  $A_{2u}^-$ && $x \hat{\bm{y}} - y\hat{\bm{x}}$ && $\hat{k}^z$ \\
  $B_{1u}^-$ && $y (3 x^2 - y^2) \hat{\bm{z}}, \, z (x^2 - y^2) \hat{\bm{y}} + 2 x y z \hat{\bm{x}}$ && $\hat{k}^x ((\hat{k}^x)^2 - 3 (\hat{k}^y)^2)$ \\
  $B_{2u}^-$ && $x (x^2 - 3 y^2) \hat{\bm{z}}, \, z (x^2 - y^2) \hat{\bm{x}} - 2 x y z \hat{\bm{y}}$ && $\hat{k}^y (3 (\hat{k}^x)^2 - (\hat{k}^y)^2)$ \\
  $E_{1u}^-$ && $\{y \hat{\bm{z}} + z \hat{\bm{y}}, z \hat{\bm{x}} + x \hat{\bm{z}}\}, \, \{y \hat{\bm{z}} - z \hat{\bm{y}}, z \hat{\bm{x}} - x \hat{\bm{z}}\}$ && $\{\hat{k}^x, \hat{k}^y\}$ \\
  $E_{2u}^-$ && $\{x \hat{\bm{y}} + y \hat{\bm{x}}, x \hat{\bm{x}} - y \hat{\bm{y}}\}$ && $\{2 \hat{k}^x \hat{k}^y \hat{k}^z, \hat{k}^z ((\hat{k}^x)^2 - (\hat{k}^y)^2)\}$ \\ \hline\hline
 \end{tabularx}
\end{table*}

\begin{table*}[tbp]
 \centering
 \caption{Basis functions in the real and momentum spaces for IRs of the $O_h$ point group~\cite{Watanabe2018}.}
 \label{tab:multipole_basis_Oh}
 \begin{tabularx}{\linewidth}{lXcXc} \hline\hline
  IR ($\Gamma$) && Basis in real space && Basis in momentum space \\ \hline
  \multicolumn{5}{c}{(EE)} \\
  $A_{1g}^+$ && $x^2 + y^2 + z^2$ && \\
  $A_{2g}^+$ && $(x^2 - y^2) (y^2 - z^2) (z^2 - x^2)$ && \\
  $E_{g}^+$  && $\{2 z^2 - x^2 - y^2, x^2 - y^2\}$ && \multicolumn{1}{c}{$(\bm{r} \rightarrow \hat{\bm{k}})$} \\
  $T_{1g}^+$ && $\{y z (y^2 - z^2), z x (z^2- x^2), x y (x^2 - y^2)\}$ && \\
  $T_{2g}^+$ && $\{y z, z x, x y\}$ && \\
  \\
  \multicolumn{5}{c}{(EM)} \\
  $A_{1g}^-$ && $y z (y^2 - z^2) \hat{\bm{x}} + z x (z^2 - x^2) \hat{\bm{y}} + x y (x^2 - y^2) \hat{\bm{z}}$ && \\
  $A_{2g}^-$ && $y z \hat{\bm{x}} + z x \hat{\bm{y}} + x y \hat{\bm{z}}$ && \\
  $E_{g}^-$  && $\{y z \hat{\bm{x}} - z x \hat{\bm{y}}, 2 x y \hat{\bm{z}} - y z \hat{\bm{x}} - z x \hat{\bm{y}}\}$ && \\
  $T_{1g}^-$ && $\{\hat{\bm{x}}, \hat{\bm{y}}, \hat{\bm{z}}\}$ && \multicolumn{1}{c}{$(\bm{r} \rightarrow \hat{\bm{k}})$} \\
  $T_{2g}^-$ && $\{x y \hat{\bm{y}} - z x \hat{\bm{z}}, y z \hat{\bm{z}} - x y \hat{\bm{x}}, z x \hat{\bm{x}} - y z \hat{\bm{y}}\},$ && \\
             && $\{(y^2 - z^2) \hat{\bm{x}}, (z^2 - x^2) \hat{\bm{y}}, (x^2 - y^2) \hat{\bm{z}}\}$ && \\
  \\
  \multicolumn{5}{c}{(OE)} \\
  $A_{1u}^+$ && $x y z (x^2 - y^2) (y^2 - z^2) (z^2 - x^2)$ && $\hat{k}^x \hat{\bm{x}} + \hat{k}^y \hat{\bm{y}} + \hat{k}^z \hat{\bm{z}}$ \\
  $A_{2u}^+$ && $x y z$ && $\hat{k}^x ((\hat{k}^y)^2 - (\hat{k}^z)^2) \hat{\bm{x}} + \hat{k}^y ((\hat{k}^z)^2 - (\hat{k}^x)^2) \hat{\bm{y}} + \hat{k}^z ((\hat{k}^x)^2 - (\hat{k}^y)^2) \hat{\bm{z}}$ \\
  $E_u^+$    && $\{x y z (2 z^2 - x^2 - y^2 ), x y z (x^2 - y^2)\}$ && $\{\hat{k}^x \hat{\bm{x}} - \hat{k}^y \hat{\bm{y}}, 2 \hat{k}^z \hat{\bm{z}} - \hat{k}^x \hat{\bm{x}} - \hat{k}^y \hat{\bm{y}}\}$ \\
  $T_{1u}^+$ && $\{x, y, z\}$ && \multirow{1}{*}{$\{\hat{k}^y \hat{\bm{z}} - \hat{k}^z \hat{\bm{y}}, \hat{k}^z \hat{\bm{x}} - \hat{k}^x\hat{\bm{z}}, \hat{k}^x \hat{\bm{y}} - \hat{k}^y\hat{\bm{x}}\}$} \\
  $T_{2u}^+$ && $\{x (y^2 - z^2), y (z^2 - x^2), z (x^2 - y^2)\}$ && $\{\hat{k}^y \hat{\bm{z}} + \hat{k}^z \hat{\bm{y}}, \hat{k}^z \hat{\bm{x}} + \hat{k}^x \hat{\bm{z}}, \hat{k}^x \hat{\bm{y}} + \hat{k}^y \hat{\bm{x}}\}$ \\
  \\
  \multicolumn{5}{c}{(OM)} \\
  $A_{1u}^-$ && $x \hat{\bm{x}} + y\hat{\bm{y}} + z \hat{\bm{z}}$ && $\hat{k}^x \hat{k}^y \hat{k}^z ((\hat{k}^x)^2 - (\hat{k}^y)^2) ((\hat{k}^y)^2 - (\hat{k}^z)^2) ((\hat{k}^z)^2 - (\hat{k}^x)^2)$ \\
  $A_{2u}^-$ && $x (y^2 - z^2) \hat{\bm{x}} + y (z^2 - x^2) \hat{\bm{y}} + z (x^2 - y^2) \hat{\bm{z}}$ && $\hat{k}^x \hat{k}^y \hat{k}^z$ \\
  $E_{u}^-$  && $\{x \hat{\bm{x}} - y \hat{\bm{y}}, 2 z \hat{\bm{z}} - x \hat{\bm{x}} - y \hat{\bm{y}}\}$ && $\{\sqrt{3} \hat{k}^x \hat{k}^y \hat{k}^z ((\hat{k}^x)^2 - (\hat{k}^y)^2), \hat{k}^x \hat{k}^y \hat{k}^z (2 (\hat{k}^z)^2 - (\hat{k}^x)^2 - (\hat{k}^y)^2)\}$ \\
  $T_{1u}^-$ && $\{y \hat{\bm{z}} - z \hat{\bm{y}}, z \hat{\bm{x}} - x \hat{\bm{z}}, x \hat{\bm{y}} - y \hat{\bm{x}}\}$ && $\{\hat{k}^x, \hat{k}^y, \hat{k}^z\}$ \\
  $T_{2u}^-$ && $\{y \hat{\bm{z}} + z \hat{\bm{y}}, z \hat{\bm{x}} + x \hat{\bm{z}}, x \hat{\bm{y}} + y \hat{\bm{x}}\}$ && $\{\hat{k}^x ((\hat{k}^y)^2 - (\hat{k}^z)^2), \hat{k}^y ((\hat{k}^z)^2 - (\hat{k}^x)^2), \hat{k}^z ((\hat{k}^x)^2 - (\hat{k}^y)^2)\}$ \\ \hline\hline
 \end{tabularx}
\end{table*}

\section{Direct product tables}
\label{ap:direct_product}
In Tables~\ref{tab:vertex_D4h}--\ref{tab:vertex_Oh}, direct products of representations are used for classification of the channel of the vertex functions.
We explicitly show the direct products for each point group in Table~\ref{tab:direct_product}.

\begin{table*}
 \caption{Direct product tables for (a) $D_{4h}$, (b) $D_{6h}$, and (c) $O_h$ point groups.}
 \label{tab:direct_product}
 \begin{tabularx}{\linewidth}{lXcXcXcXcXcXcXcXcXcXc} \hline\hline
  \multicolumn{21}{c}{(a) Tetragonal ($D_{4h}$)} \\ \hline
  $\Gamma$ && $A_{1g}$ &&	$A_{2g}$ &&	$B_{1g}$ &&	$B_{2g}$ && $E_g$ &&	$A_{1u}$ &&	$A_{2u}$ &&	$B_{1u}$ &&	$B_{2u}$ &&	$E_u$ \\ \hline
  $\Gamma \times A_{2g}$ && $A_{2g}$ && $A_{1g}$	&& $B_{2g}$	&& $B_{1g}$	&& $E_g$	&& $A_{2u}$	&& $A_{1u}$	&& $B_{2u}$	&& $B_{1u}$	&& $E_u$ \\
  $\Gamma \times E_g$ && $E_g$	&& $E_g$	&& $E_g$	&& $E_g$	&& $A_{1g} + A_{2g} + B_{1g} + B_{2g}$	&& $E_u$	&& $E_u$	&& $E_u$	&& $E_u$	&& $A_{1u} + A_{2u} + B_{1u} + B_{2u}$ \\
 \end{tabularx}
 \\[5mm]
 \begin{tabularx}{\linewidth}{lXcXcXcXcXcXcXcXcXcXcXcXc}
  \multicolumn{25}{c}{(b) Hexagonal ($D_{6h}$)} \\ \hline
  $\Gamma$ && $A_{1g}$ &&	$A_{2g}$ &&	$B_{1g}$ &&	$B_{2g}$ && $E_{1g}$ && $E_{2g}$ &&	$A_{1u}$ &&	$A_{2u}$ &&	$B_{1u}$ &&	$B_{2u}$ && $E_{1u}$ && $E_{2u}$ \\ \hline
  $\Gamma \times A_{2g}$ && $A_{2g}$ && $A_{1g}$	&& $B_{2g}$	&& $B_{1g}$	&& $E_{1g}$ && $E_{2g}$	&& $A_{2u}$	&& $A_{1u}$	&& $B_{2u}$	&& $B_{1u}$	&& $E_{1u}$ && $E_{2u}$ \\
  $\Gamma \times E_{1g}$ && $E_{1g}$	&& $E_{1g}$	&& $E_{2g}$	&& $E_{2g}$	&& $A_{1g} + A_{2g} + E_{2g}$	&& $B_{1g} + B_{2g} + E_{1g}$	&& $E_{1u}$	&& $E_{1u}$	&& $E_{2u}$	&& $E_{2u}$	&& $A_{1u} + A_{2u} + E_{2u}$	&& $B_{1u} + B_{2u} + E_{1u}$ \\
 \end{tabularx}
 \\[5mm]
 \begin{tabularx}{\linewidth}{lXcXcXcXC{15mm}XC{15mm}XcXcXcXC{15mm}XC{15mm}}
  \multicolumn{21}{c}{(c) Cubic ($O_h$)} \\ \hline
  $\Gamma$ && $A_{1g}$ &&	$A_{2g}$ &&	$E_g$ && $T_{1g}$ && $T_{2g}$ &&	$A_{1u}$ &&	$A_{2u}$ &&	$E_u$ && $T_{1u}$ && $T_{2u}$ \\ \hline
  $\Gamma \times T_{1g}$ && $T_{1g}$	&& $T_{2g}$ && $T_{1g} + T_{2g}$	&& $A_{1g} + E_g + T_{1g} + T_{2g}$	&& $A_{2g} + E_g + T_{1g} + T_{2g}$ && $T_{1u}$	&& $T_{2u}$ && $T_{1u} + T_{2u}$	&& $A_{1u} + E_u + T_{1u} + T_{2u}$	&& $A_{2u} + E_u + T_{1u} + T_{2u}$ \\ \hline\hline
 \end{tabularx}
\end{table*}

\section{Vector spherical harmonics}
\label{ap:vector_spherical_harmonics}
In this appendix, we introduce vector spherical harmonics~\cite{Varshalovich}, which are useful for the formulation of the linearized gap equation for spin-triplet superconductivity (Appendix~\ref{ap:linearized_gap_eq}).

\subsection{Definition}
Vector spherical harmonics are defined by
\begin{equation}
 \bm{Y}_{J M}^{L}(\hat{\bm{k}}) \equiv \sum_{m = - L}^{L} \sum_{\mu = - 1}^{1} C_{L m 1 \mu}^{J M} Y_{L m}(\hat{\bm{k}}) \bm{e}_{\mu},
 \label{eq:vector_spherical_harmonics}
\end{equation}
where $Y_{L m}(\hat{\bm{k}})$ are (usual) spherical harmonics,\footnote{For the later discussion in Appendix~\ref{ap:linearized_gap_eq}, we define (vector) spherical harmonics in momentum ($\hat{\bm{k}}$) space, not in real ($\hat{\bm{r}}$) space.} and $C_{L m 1 \mu}^{J M}$ are Clebsch-Gordan coefficients.
$\bm{e}_\mu$ ($\mu = -1, 0, 1$) are covariant spherical basis vectors (spin functions for $S = 1$),
\begin{equation}
 \bm{e}_{-1} = \frac{1}{\sqrt{2}} (\hat{\bm{x}} - i\hat{\bm{y}}), \, \bm{e}_{0} = \hat{\bm{z}}, \, \bm{e}_{1} = - \frac{1}{\sqrt{2}} (\hat{\bm{x}} + i\hat{\bm{y}}),
\end{equation}
which have an orthonormal property $\bm{e}_{\mu_1} \cdot \bm{e}_{\mu_2}^* = \delta_{\mu_1 \mu_2}$.

As simple examples, the vector spherical harmonics for $L = 0, 1$ are listed here:
\begin{subequations}
 \begin{align}
  \bm{Y}_{1 M}^{0}(\hat{\bm{k}}) &= \frac{1}{\sqrt{4\pi}} \bm{e}_M \quad (M = -1, 0, 1), \displaybreak[2] \\
  \bm{Y}_{0 0}^{1}(\hat{\bm{k}}) &= - \frac{1}{\sqrt{4\pi}} (\hat{k}^x \hat{\bm{x}} + \hat{k}^y \hat{\bm{y}} + \hat{k}^z \hat{\bm{z}}), \displaybreak[2] \\
  \bm{Y}_{1 -1}^{1}(\hat{\bm{k}}) &= \frac{1}{\sqrt{4\pi}} \frac{\sqrt{3}}{2} \{\hat{k}^z (\hat{\bm{x}} - i\hat{\bm{y}}) - (\hat{k}^x - i\hat{k}^y) \hat{\bm{z}}\}, \\
  \bm{Y}_{1 0}^{1}(\hat{\bm{k}}) &= \frac{i}{\sqrt{4\pi}} \sqrt{\frac{3}{2}} (\hat{k}^x \hat{\bm{y}} - \hat{k}^y \hat{\bm{x}}), \\
  \bm{Y}_{1 1}^{1}(\hat{\bm{k}}) &= \frac{1}{\sqrt{4\pi}} \frac{\sqrt{3}}{2} \{\hat{k}^z (\hat{\bm{x}} + i\hat{\bm{y}}) - (\hat{k}^x + i\hat{k}^y) \hat{\bm{z}}\}, \displaybreak[2] \\
  \bm{Y}_{2 -2}^{1}(\hat{\bm{k}}) &= \frac{1}{\sqrt{4\pi}} \frac{\sqrt{3}}{2} (\hat{k}^x - i\hat{k}^y) (\hat{\bm{x}} - i\hat{\bm{y}}), \\
  \bm{Y}_{2 -1}^{1}(\hat{\bm{k}}) &= \frac{1}{\sqrt{4\pi}} \frac{\sqrt{3}}{2} \{\hat{k}^z (\hat{\bm{x}} - i\hat{\bm{y}}) + (\hat{k}^x - i\hat{k}^y) \hat{\bm{z}}\}, \\
  \bm{Y}_{2 0}^{1}(\hat{\bm{k}}) &= \frac{1}{\sqrt{4\pi}} \frac{1}{\sqrt{2}} (2 \hat{k}^z \hat{\bm{z}} - \hat{k}^x \hat{\bm{x}} - \hat{k}^y \hat{\bm{y}}), \\
  \bm{Y}_{2 1}^{1}(\hat{\bm{k}}) &= - \frac{1}{\sqrt{4\pi}} \frac{\sqrt{3}}{2} \{\hat{k}^z (\hat{\bm{x}} + i\hat{\bm{y}}) + (\hat{k}^x + i\hat{k}^y) \hat{\bm{z}}\}, \\
  \bm{Y}_{2 2}^{1}(\hat{\bm{k}}) &= \frac{1}{\sqrt{4\pi}} \frac{\sqrt{3}}{2} (\hat{k}^x + i\hat{k}^y) (\hat{\bm{x}} + i\hat{\bm{y}}).
 \end{align}
\end{subequations}

\subsection{Properties}
The vector spherical harmonics [Eq.~\eqref{eq:vector_spherical_harmonics}] satisfy the following orthonormal property,
\begin{equation}
 \int d\Omega_{\hat{\bm{k}}} \, \bm{Y}_{J_1 M_1}^{L_1}(\hat{\bm{k}}) \cdot \bm{Y}_{J_2 M_2}^{L_2}(\hat{\bm{k}})^* = \delta_{J_1 J_2} \delta_{M_1 M_2} \delta_{L_1 L_2},
\end{equation}
where $\Omega_{\hat{\bm{k}}}$ is a solid angle of the normalized vector $\hat{\bm{k}}$.
This equality is proved as follows:
\begin{align}
 (\text{LHS}) &= \sum_{m_1 = - L_1}^{L_1} \sum_{\mu_1 = - 1}^{1} \sum_{m_2 = - L_2}^{L_2} \sum_{\mu_2 = - 1}^{1} C_{L_1 m_1 1 \mu_1}^{J_1 M_1} C_{L_2 m_2 1 \mu_2}^{J_2 M_2} \notag \\
 & \quad \times \underbrace{\int d\Omega_{\hat{\bm{k}}} \, Y_{L_1 m_1}(\hat{\bm{k}}) Y_{L_2 m_2}^*(\hat{\bm{k}})}_{\delta_{L_1 L_2} \delta_{m_1 m_2}} \underbrace{\bm{e}_{\mu_1} \cdot \bm{e}_{\mu_2}^*}_{\delta_{\mu_1 \mu_2}} \notag \displaybreak[2] \\
 &= \sum_{m_1 = - L_1}^{L_1} \sum_{\mu_1 = - 1}^{1} C_{L_1 m_1 1 \mu_1}^{J_1 M_1} C_{L_1 m_1 1 \mu_1}^{J_2 M_2} \delta_{L_1 L_2} \notag \\
 &= (\text{RHS}),
\end{align}
where we use the orthogonality relation of the Clebsch-Gordan coefficients in the final equality.

In a similar way, the following ``orthogonality'' is easily derived:
\begin{align}
 &\int d\Omega_{\hat{\bm{k}}} \sum_{\alpha \beta} \{\bm{Y}_{J_1 M_1}^{L_1}(\hat{\bm{k}}) \cdot \bm{\sigma} i\sigma^y\}_{\alpha \beta} \{\bm{Y}_{J_2 M_2}^{L_2}(\hat{\bm{k}}) \cdot \bm{\sigma} i\sigma^y\}^\dagger_{\alpha \beta} \notag \\
 & \qquad = 2 \delta_{J_1 J_2} \delta_{M_1 M_2} \delta_{L_1 L_2},
 \label{eq:vector_spherical_harmonics_orthogonality}
\end{align}
where the factor $2$ on the RHS stems from the trace of the $2 \times 2$ matrix,
\begin{equation}
 \sum_{\alpha \beta} (\bm{e}_{\mu_1} \cdot \bm{\sigma} i\sigma^y)_{\alpha \beta} (\bm{e}_{\mu_2} \cdot \bm{\sigma} i\sigma^y)^\dagger_{\alpha \beta} = 2 \delta_{\mu_1 \mu_2}.
\end{equation}
Equation~\eqref{eq:vector_spherical_harmonics_orthogonality} is a useful relation for the formulation of the linearized gap equation for spin-triplet superconductivity (Appendix~\ref{ap:linearized_gap_eq_triplet}).

\section{Linearized gap equation}
\label{ap:linearized_gap_eq}
Here we introduce a basic formulation of a linearized gap equation for spin-singlet (even-parity) and spin-triplet (odd-parity) superconductivity.
Note that spin-space isotropy is not assumed, unlike the usual ``textbook-style'' formulation where spin-singlet and spin-triplet cases are not distinguished.
Instead, we postulate that the energy dispersion is quadratic (isotropic): $\xi_{\bm{k}} = \xi_{k} = \frac{k^2}{2m} - \mu$.

In a single-band problem, the Hamiltonian is
\begin{align}
 H &= \sum_{\bm{k}} \sum_{\alpha} \xi_{k} c_{\bm{k} \alpha}^\dagger c_{\bm{k} \alpha} \notag \\
 & \quad + \frac{1}{2} \sum_{\bm{k}, \bm{k}'} \sum_{\alpha \beta \gamma \delta} V_{\alpha \beta \gamma \delta}(\bm{k}, \bm{k}') c_{-\bm{k} \alpha}^\dagger c_{\bm{k} \beta}^\dagger c_{\bm{k}' \gamma} c_{-\bm{k}' \delta},
\end{align}
where we assume the existence of spatial inversion and TRS.
Using a mean-field approximation, the effective Hamiltonian is given by
\begin{align}
 H_{\text{MF}} &= \sum_{\alpha} \xi_{k} c_{\bm{k} \alpha}^\dagger c_{\bm{k} \alpha} \notag \\
 & \quad + \frac{1}{2} \sum_{\bm{k}} \sum_{\alpha \beta} \left\{ \Delta_{\alpha \beta}(\bm{k}) c_{\bm{k} \alpha}^\dagger c_{-\bm{k} \beta}^\dagger + \text{H.c.} \right\} + \text{const.},
\end{align}
where
\begin{equation}
 \Delta_{\alpha \beta}(\bm{k}) = - \sum_{\bm{k}'} \sum_{\gamma \delta} V_{\beta \alpha \gamma \delta}(\bm{k}, \bm{k}') \braket{c_{\bm{k}' \gamma} c_{-\bm{k}' \delta}}.
\end{equation}
Through a Bogoliubov transformation, we obtain the self-consistent gap equation,
\begin{equation}
 \Delta_{\alpha \beta}(\bm{k}) = - \sum_{\bm{k}'} \sum_{\gamma \delta} V_{\beta \alpha \gamma \delta}(\bm{k}, \bm{k}') \frac{\Delta_{\gamma \delta}(\bm{k}')}{2 E_{\bm{k}'}} \tanh\left(\frac{E_{\bm{k}'}}{2 T}\right),
\end{equation}
with
\begin{equation}
 E_{\bm{k}} = \sqrt{\xi_{k}^2 + \frac{1}{2} \Tr[\Delta(\bm{k})\Delta^\dagger(\bm{k})]},
\end{equation}
where a unitary order parameter is assumed.
Just below the transition temperature $T_{\text{c}}$, $\Delta(\bm{k})$ is negligibly small so that $E_{\bm{k}} = \xi_{k}$.
Therefore, the following linearized gap equation is obtained:
\begin{equation}
 \Delta_{\alpha \beta}(\bm{k}) = - \sum_{\bm{k}'} \sum_{\gamma \delta} V_{\beta \alpha \gamma \delta}(\bm{k}, \bm{k}') \frac{\Delta_{\gamma \delta}(\bm{k}')}{2 \xi_{k'}} \tanh\left(\frac{\xi_{k'}}{2 T_{\text{c}}}\right).
 \label{eq:linearized_gap_eq}
\end{equation}

\subsection{Spin-singlet superconductivity}
\label{ap:linearized_gap_eq_singlet}
Next, we solve the linearized gap equation~\eqref{eq:linearized_gap_eq} for a spin-singlet interaction channel.
In the channel, the interaction vertex can be expanded by spherical harmonics,
\begin{align}
 & V^{\text{singlet}}_{\alpha \beta \gamma \delta}(\bm{k}, \bm{k}') \notag \\
 &= 4\pi \sum_{L = \text{even}} V_{L}(k, k') \sum_{m = -L}^{L} Y_{L m}(\hat{\bm{k}}) Y_{L m}^*(\hat{\bm{k}}') \notag \\
 & \quad \times \frac{1}{2} (i\sigma^y)_{\alpha \beta} (i\sigma^y)^\dagger_{\gamma \delta},
\end{align}
where the product of two spherical harmonics $Y_{L m}(\hat{\bm{k}}) Y_{L m}^*(\hat{\bm{k}}')$ appears with the same $L$ and $m$ because isotropic systems are assumed.
Note that a cross term $Y_{L m}(\hat{\bm{k}}) Y_{L' m'}^*(\hat{\bm{k}}')$ ($L \neq L'$ or $m \neq m'$) is allowed in crystalline systems [e.g. Eq.~\eqref{eq:interaction_vertex_D4h_A2up_A1g}]; in such a case, we appropriately take into account cross terms and derive simultaneous equations.
Below we show solutions for isotropic systems.
The superconducting order parameter is also expanded by spherical harmonics~\cite{Sigrist-Ueda},
\begin{equation}
 \Delta^{\text{singlet}}(\bm{k}) = \sum_{L = \text{even}} \sum_{m = -L}^{L} c_{L m} Y_{L m}(\hat{\bm{k}}) i\sigma^y.
\end{equation}

Using the orthonormal property of spherical harmonics, Eq.~\eqref{eq:linearized_gap_eq} is simplified as
\begin{equation}
 1 = - \frac{\Omega}{2\pi^2} \int_{0}^{\infty} dk' \, k'^2 V_{L}(k, k') \frac{1}{2\xi_{k'}} \tanh\left(\frac{\xi_{k'}}{2 T_{\text{c}}}\right),
\end{equation}
where $\Omega$ is a volume.
Now we assume that $V_{L}(k, k')$ is finite nearby the Fermi surface; there exists a cutoff energy $\omega_c$ such that
\begin{equation}
 V_{L}(k, k') =
 \begin{cases}
  - V_{L} \ (\text{const.}) & |\xi_{k}|, |\xi_{k'}| \leq \omega_c, \\
  0 & |\xi_{k}|, |\xi_{k'}| > \omega_c.
 \end{cases}
\end{equation}
Then, the linearized gap equation is
\begin{equation}
 1 = N(0) V_{L} \int_{0}^{\omega_c} d\xi \, \frac{1}{\xi} \tanh\left(\frac{\xi}{2 T_{\text{c}}}\right) \qquad (L = \text{even}),
 \label{eq:linearized_gap_eq_singlet}
\end{equation}
where $N(0)$ is a density of states at the Fermi level.
Therefore, the solution (order parameter) of Eq.~\eqref{eq:linearized_gap_eq_singlet} is labeled by $L$: $\Delta_{L}(\bm{k}) = \sum_{m = -L}^{L} c_{L m} Y_{L m}(\hat{\bm{k}}) i\sigma^y$.
For each $L$, the transition temperature is given by
\begin{equation}
 T_{\text{c}}^{(L)} = 1.14 \omega_c \exp\left(- \frac{1}{N(0) V_{L}}\right).
\end{equation}

\subsection{Spin-triplet superconductivity}
\label{ap:linearized_gap_eq_triplet}
Let us move on to the spin-triplet case.
The interaction vertex and the superconducting order parameter in the spin-triplet channel can be expanded by vector spherical harmonics,
\begin{align}
 & V^{\text{triplet}}_{\alpha \beta \gamma \delta}(\bm{k}, \bm{k}') \notag \\
 &= 4\pi \sum_{J} \sum_{L = \text{odd}} V_{J L}(k, k') \notag \\
 & \quad \times \frac{1}{2} \sum_{M = -J}^{J} \{\bm{Y}_{J M}^{L}(\hat{\bm{k}}) \cdot \bm{\sigma} i\sigma^y\}_{\alpha \beta} \{\bm{Y}_{J M}^{L}(\hat{\bm{k}}') \cdot \bm{\sigma} i\sigma^y\}^\dagger_{\gamma \delta}, \displaybreak[2] \\
 & \Delta^{\text{triplet}}(\bm{k}) = \sum_{J} \sum_{L = \text{odd}} \sum_{M = -J}^{J} c_{J M}^{L} \bm{Y}_{J M}^{L}(\hat{\bm{k}}) \cdot \bm{\sigma} i\sigma^y.
\end{align}

Using the orthonormal property of vector spherical harmonics (Appendix~\ref{ap:vector_spherical_harmonics}), Eq.~\eqref{eq:linearized_gap_eq} is
\begin{equation}
 1 = - \frac{\Omega}{2\pi^2} \int_{0}^{\infty} dk' \, k'^2 V_{J L}(k, k') \frac{1}{2\xi_{k'}} \tanh\left(\frac{\xi_{k'}}{2 T_{\text{c}}}\right).
\end{equation}
Also we assume that $V_{L}(k, k')$ is finite nearby the Fermi surface,
\begin{equation}
 V_{J L}(k, k') =
 \begin{cases}
  - V_{J L} \ (\text{const.}) & |\xi_{k}|, |\xi_{k'}| \leq \omega_c, \\
  0 & |\xi_{k}|, |\xi_{k'}| > \omega_c.
 \end{cases}
\end{equation}
Then, the linearized gap equation is simplified as
\begin{equation}
 1 = N(0) V_{J L} \int_{0}^{\omega_c} d\xi \, \frac{1}{\xi} \tanh\left(\frac{\xi}{2 T_{\text{c}}}\right) \qquad (L = \text{odd}).
 \label{eq:linearized_gap_eq_triplet}
\end{equation}
Therefore, the solution (order parameter) of Eq.~\eqref{eq:linearized_gap_eq_triplet} is labeled by $J$ and $L$: $\Delta_{J L}(\bm{k}) = \sum_{M = -L}^{L} c_{J M}^{L} \bm{Y}_{J M}^{L}(\hat{\bm{k}}) \cdot \bm{\sigma} i\sigma^y$.
For each $(J, L)$, the transition temperature is given by
\begin{equation}
 T_{\text{c}}^{(J L)} = 1.14 \omega_c \exp\left(- \frac{1}{N(0) V_{J L}}\right).
\end{equation}

\end{document}